\documentclass[graybox]{svmult}

\usepackage{graphicx}

\usepackage{amsmath}
\usepackage{color}

\usepackage[latin1]{inputenc}
\usepackage{amsfonts}
\usepackage{tikz} 
\usetikzlibrary{matrix}
\usepackage[english]{babel}
\usepackage[latin1]{inputenc}
\usepackage[T1]{fontenc}
\usepackage{ae}

\usepackage{url}

\allowdisplaybreaks[4]

\usepackage{color}
\usepackage{tikz}
\usetikzlibrary{matrix}
\usepackage{cite}           
\usepackage{mathptmx}       
\usepackage{helvet}         
\usepackage{courier}        
\usepackage{type1cm}        
\usepackage{makeidx}         
\usepackage{graphicx}        
\usepackage{multicol}        
\usepackage[bottom]{footmisc}

\usepackage{tikz}
\usepackage{etex, xy}
\xyoption{all}
\usetikzlibrary{matrix}

\definecolor{blaugrau}{rgb}{0.796887, 0.789075, 0.871107}

\newcounter{linectr}
\newenvironment{myEnumerate}{\begin{list}{(\arabic{linectr})}{\usecounter{linectr}
\labelwidth1ex\itemsep0ex\labelsep1ex\leftmargin2ex\parskip0.0cm\topskip0cm\partopsep0cm
\listparindent0ex}}{\end{list}}

\allowdisplaybreaks[4]

\newcounter{mmacnt}
\def\restartmma{\setcounter{mmacnt}{0}}
\restartmma \catcode`|=\active
\def|#1|{\mathrm{#1}}
\catcode`|=12
\newenvironment{mma}{
 \par
 \catcode`|=\active
 \parskip=2pt\parindent=0pt 
 \small
 \def\In##1\\{%
   \def\linebreak{\hfill\break\null\qquad}%
   \refstepcounter{mmacnt}
   \hangindent=2.5em\hangafter=0
   \leavevmode
   \llap{\tiny\sffamily In[\arabic{mmacnt}]:=\kern.5em}%
   \mathversion{bold}\scriptsize$\tt\bf\displaystyle##1$\normalsize
   \mathversion{normal}\par
 }%
 \def\Print##1\\{%
   \def\linebreak{\hfill\break}%
   \hangindent=2.5em\hangafter=0
   \leavevmode\scriptsize ##1\par}%
 \def\Out##1\\{%
   \vspace*{-0.2cm}\def\linebreak{$\hfill\break\null\hfill$}%
   \kern\abovedisplayskip\par
   \hangindent=2.5em\hangafter=0
   \leavevmode
   \llap{\tiny\sffamily Out[\arabic{mmacnt}]=\kern.5em}
   \scriptsize$\displaystyle\tt##1$\normalsize\hfill\null\par
   \kern\belowdisplayskip\vspace*{-0.3cm}
 }%
 \def\Warning##1##2\\{%
   \def\linebreak{\hfill\break}%
   \hangindent=2.5em\hangafter=0
   \leavevmode
   {\scriptsize##1 : ##2}\par}%
}{%
 \par\smallskip
}

\newcommand{\LoadP}[1]{\fcolorbox{black}{blaugrau}{
\begin{minipage}[t]{10.2cm}
\footnotesize #1
\end{minipage}}}

\newcommand{\myIn}[1]{{\sffamily In[#1]}}
\newcommand{\myOut}[1]{{\sffamily Out[#1]}}

\def\MLabel#1{{\refstepcounter{mmacnt}\label{#1}}\addtocounter{mmacnt}{-1}}

\newcommand{\MInText}[1]{\text{\mathversion{bold}\scriptsize$\tt\bf\displaystyle#1$}}

\let\set\mathbb

\def\ev{\operatorname{ev}}

\def\rE{$R$}
\def\piE{$\Pi$}

\def\sigmaSE{$\Sigma$}
\def\rpisiSE{$R\Pi\Sigma$}

\def\seqK{\mathrm{Seq}(\KK)}
\newcommand{\Shift}{{S}}

\def\AA{\set A}

\def\SubS{\set S}
\def\KK{\set K}
\def\NN{{\set Z}_{\geq0}}
\def\ZZ{\set Z}
\def\GG{\set G}

\def\FF{\set F}
\def\EE{\set E}
\def\QQ{\set Q}

\spnewtheorem{psdefinition}{Definition}[section]{\bf}{}
\spnewtheorem{pstheorem}[psdefinition]{Theorem}{\bf}{}
\spnewtheorem{psproposition}[psdefinition]{Proposition}{\bf}{}
\spnewtheorem{pscorollary}[psdefinition]{Corollary}{\bf}{}
\spnewtheorem{pslemma}[psdefinition]{Lemma}{\bf}{}
\spnewtheorem{psremark}[psdefinition]{Remark}{\bf}{}
\spnewtheorem{psexample}[psdefinition]{Example}{\bf}{}
\spnewtheorem{psconvention}[psdefinition]{Convention}{\bf}{}

\newcommand{\fct}[3]{#1:#2\to#3}

\newcommand{\dfield}[2]{(#1,#2)}
\newcommand{\const}[2]{\operatorname{const}(#1,#2)}

\makeindex             

\begin{document}

\title*{Towards a symbolic summation theory for unspecified sequences}
\author{Peter Paule and Carsten Schneider}
\institute{Peter Paule and Carsten Schneider,\\  Research Institute for Symbolic Computation (RISC)
Johannes Kepler University, Altenbergerstr.~69, 4040 Linz, Austria,
\email{ppaule@risc.jku.at, cschneid@risc.jku.at}}
%
%
\maketitle

\abstract{The article addresses the problem whether indefinite double sums involving a generic sequence can be simplified in terms of indefinite single sums. Depending on the structure of the double sum, the proposed summation machinery may provide such a simplification without exceptions. If it fails, it may suggest a more advanced simplification introducing in addition a single nested sum where the summand has to satisfy a particular constraint. More precisely, an explicitly given parameterized telescoping equation must hold. Restricting to the case that the arising unspecified sequences are specialized to the class of indefinite nested sums defined over hypergeometric, multi-basic or mixed hypergeometric products, it can be shown that this constraint is not only sufficient but also necessary.}

\section{Introduction} \label{intro}

Over recent years the second named author succeeded in developing a difference field (resp.\ ring) theory which allows to treat within a common algorithmic framework summation problems with elements from algebraically specified domains as well as problems involving concrete sequences which are analytically specified (e.g., from quantum field theory, combinatorics, number theory, and special functions). In this article we establish a new algebraic/algorithmic connection between this setting and summation problems involving generic sequences. We feel there is a high application potential for this connection. One future domain for algorithmic discovery (as described below) might be identities involving elliptic functions and modular forms.  

In the course of a project devoted to an algorithmic
revival of MacMahon's partition analysis, Andrews
and Paule showed in~\cite{PAIV} that a variant of partition
analysis can be applied also for simplification of
multiple combinatorial sums. Starting with the pioneering
work of Abramov~\cite{Abramov:71,Abramov:89a}, Gosper~\cite{Gosper:78}, Karr~\cite{Karr:81,Karr:85}, and Zeilberger~\cite{Zeilberger:90b}, significant progress has been made. In particular,
in the context of summation in difference fields and, more generally, difference rings~\cite{DR1,DR2,DR3} Schneider has 
developed substantial extensions and generalizations~\cite{Schneider:07d,FastAlgorithm1,FastAlgorithm2,FastAlgorithm3}
of Karr's seminal work. Owing to such an algorithmic
machinery, the summation problems treated in~\cite{PAIV} can nowadays be done in a jiffy with Schneider's \texttt{Sigma} package~\cite{Sigma1}.

Nevertheless, the present article connects to~\cite{PAIV}
in various ways. First, it also considers a class of
summation identities related to the celebrated 
Calkin sum which is the case $\ell=3$ of
\[
C_\ell(n):=\sum_{k=0}^n\left(
\sum_{j=0}^k \binom{n}{j}\right)^\ell.
\]
More generally, we will focus also on the truncated versions 
\[
C_\ell(a,n):=\sum_{k=0}^a\left(\sum_{j=0}^k \binom{n}{j}\right)^\ell.
\]
And second, similarly to~\cite{PAIV} presenting a 
``non-standard'' variation of the method of partition analysis, we present ``non-standard'' variations of difference field summation techniques.

The first ``non-standard'' ingredient is the aspect
of ``generic'' summation in difference fields and rings. First
pioneering steps in this direction were made by
Kauers and Schneider; see~\cite{KS:06a,KS:06b}. 

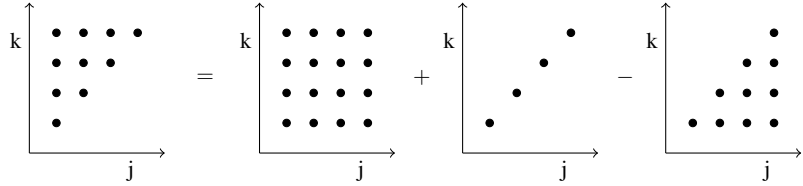
\begin{figure}
\begin{tikzpicture}[x=0.9cm]
    \coordinate (o1) at (0,0);
    \coordinate (x1) at (2,0);
    \draw[->] (o1) -- (x1) node [near end,below] {j};
    \coordinate (y1) at (0,2);
    \draw[->] (o1) -- (y1) node [near end,left] {k};
   \foreach \y in {1.6}{
        \node[draw,circle,inner sep=1pt,fill] at (1.6,\y) {};
    }
    \foreach \y in {1.2,1.6}{
        \node[draw,circle,inner sep=1pt,fill] at (1.2,\y) {};
    }
    \foreach \y in {0.8,1.2,1.6}{
        \node[draw,circle,inner sep=1pt,fill] at (0.8,\y) {};
    }
    \foreach \y in {0.4,0.8,1.2,1.6}{
        \node[draw,circle,inner sep=1pt,fill] at (0.4,\y) {};
    }

    \draw (2.6,1) node[font=\small] {$=$};

    \coordinate (o2) at (3.4,0);
    \coordinate (x2) at (5.4,0);
    \draw[->] (o2) -- (x2) node [near end,below] {j};
    \coordinate (y2) at (3.4,2);
    \draw[->] (o2) -- (y2) node [near end,left] {k};
    \foreach \x in {3.8,4.2,...,5.0}{
      \foreach \y in {0.4,0.8,...,2}{
        \node[draw,circle,inner sep=1pt,fill] at (\x,\y) {};
      }
    }

    \draw (5.8,1) node[font=\small] {$+$};

    \coordinate (o3) at (6.4,0);
    \coordinate (x3) at (8.4,0);
    \draw[->] (o3) -- (x3) node [near end,below] {j};
    \coordinate (y3) at (6.4,2);
    \draw[->] (o3) -- (y3) node [near end,left] {k};

     \foreach \y in {0.4,0.8,1.2,1.6}{
        \node[draw,circle,inner sep=1pt,fill] at (6.4+\y,\y) {};
    }

    \coordinate (o4) at (9.4,0);
    \coordinate (x4) at (11.4,0);
    \draw[->] (o4) -- (x4) node [near end,below] {j};
    \coordinate (y4) at (9.4,2);
    \draw[->] (o4) -- (y4) node [near end,left] {k};
 
    \draw (8.8,1) node[font=\small] {$-$};

    \foreach \y in {0.4}{
        \node[draw,circle,inner sep=1pt,fill] at (9.8,\y) {};
    }
    \foreach \y in {0.4,0.8}{
        \node[draw,circle,inner sep=1pt,fill] at (10.2,\y) {};
    }
    \foreach \y in {0.4,0.8,1.2}{
        \node[draw,circle,inner sep=1pt,fill] at (10.6,\y) {};
    }
    \foreach \y in {0.4,0.8,1.2,1.6}{
        \node[draw,circle,inner sep=1pt,fill] at (11.0,\y) {};
    }

\end{tikzpicture}
\caption{\label{Fig:MultiplySums} Summing over a rectangular grid in two different
ways.}
\end{figure}

To illustrate the generic aspect, consider the problem
of simplifying the sums
\[
C_1(a,n)=\sum_{k=0}^a \sum_{j=0}^k \binom{n}{j}\quad\text{ and }\quad C_1(n)=\sum_{k=0}^n \sum_{j=0}^k \binom{n}{j}.
\]
A rewriting of $C_1(a,n)$ is obtained by specializing
$Y_k=1$ and $X_j= \binom{n}{j}$
in the generic summation relation
\begin{align}\label{square rel}
\sum_{k=0}^a \left(\sum_{j=0}^k X_j\right)Y_k& =
 \left(\sum_{k=0}^a Y_k \right)
\left( \sum_{j=0}^a X_j \right)+ \sum_{k=0}^a Y_kX_k 
- \sum_{k=0}^a X_k\left( \sum_{j=0}^k  Y_j \right).
\end{align}
Pictorially, \eqref{square rel} corresponds to
summing over a square shaped grid in two different
ways; see Fig~\ref{Fig:MultiplySums}.

Specializing~\eqref{square rel} as proposed results in
\begin{align*}
C_1(a,n)=&(a+1)\sum_{j=0}^a \binom{n}{j}+\sum_{k=0}^a \binom{n}{k}-\sum_{k=0}^a \binom{n}{k}(k+1)\\
=&(a+1) \sum_{k=0}^a \binom{n}{k}
- \sum_{k=0}^a k \binom{n}{k}.
\end{align*}
This means that the application of~\eqref{square rel}
indeed results in a simplification: the original
double sum is expressed in terms of single sums. Specializing $a=n$
the single sums in turn simplify further by the
binomial theorem:
\[ 
\sum_{k=0}^n k\, \binom{n}{k} 
= n\, \sum_{k=1}^n \binom{n-1}{k-1} 
 = n\, \sum_{k=0}^{n-1} \binom{n-1}{k} 
 = n\, 2^{n-1}.
\]
This yields
\[
C_1(n)=C_1(n,n)=(n+1) 2^n - n\, 2^{n-1}= 2^{n-1} (n+2).
\]

We remark that the generic formula \eqref{square rel}
can be obtained with the \texttt{Sigma} package\footnote{Freely available with password request at\\ \texttt{http://www.risc.jku.at/research/combinat/software/Sigma/}.}:

\begin{mma}
\In << Sigma.m \\
\Print \LoadP{Sigma - A summation package by Carsten Schneider
\copyright\ RISC-JKU}\\
\end{mma}

\begin{mma}\MLabel{MMA:mySum1}
 \In mySum1=SigmaSum[Y[k]SigmaSum[X[j],{j,0,k}],{k,0,a}]\\
 \Out \sum_{k=0}^a \Bigg(
        \sum_{j=0}^k X[j]\Bigg) Y[k]\\
\end{mma}

\begin{mma}\MLabel{MMA:XYExp1}
 \In res1=SigmaReduce[mySum1,XList\to\{X,Y\},XWeight\to\{2,1\},\newline
 \hspace*{2cm} SimplifyByExt\to MinDepth, SimpleSumRepresentation\to True]\\
 \Out -\sum_{i=0}^a \Bigg(
        \sum_{j=0}^i Y[j]\Bigg) X[i]
+\Bigg(
        \sum_{i=0}^a X[i]\Bigg) 
\Bigg(\sum_{i=0}^a Y[i]\Bigg)
+
\sum_{i=0}^a X[i] Y[i]\\
\end{mma}

\begin{psremark}\label{Remark:XYCalkin1}
 Applying \texttt{SigmaReduce} with the option \MInText{XList\to\{X,Y\}} one activates the summation algorithms given in~\cite{KS:06a,FastAlgorithm1} by telling \texttt{Sigma} that $X[j](=X_j)$ and $Y[k](=Y_k)$ are generic sequences. With the option \texttt{SimplifyByExt$\to$MinDepth} the underlying algorithms try to simplify the sum~\myIn{\ref{MMA:mySum1}} so that the nested depth (i.e., the number of nested sum quantifiers) is minimized. Moreover, the option \texttt{SimpleSumRepresentation$\to$True} implies that the found sum representations have only denominators, if possible, that are linear. For this particular instance, the underlying algorithm would detect that the input expression cannot be simplified further if $X$ and $Y$ are considered as equally complicated. However, using in addition the option \MInText{XWeight\to\{2,1\}} one tells \texttt{Sigma} that $X[k]$ is counted as a more nested expression than $Y[k]$. This extra information will finally produce the output given in~\myOut{\ref{MMA:XYExp1}} by introducing the sum $\sum_{i=0}^a \big(
        \sum_{j=0}^i Y[j]\big) X[i]$ which is considered as simpler than the sum~\myIn{\ref{MMA:mySum1}}.
\end{psremark}

Next we apply the same strategy to
\[
C_2(a,n)=
\sum_{k=0}^a \left(\sum_{j=0}^k \binom{n}{j}\right)^2
\quad\text{ and }\quad C_2(n)=
\sum_{k=0}^n \left(\sum_{j=0}^k \binom{n}{j}\right)^2.
\]
A generic formula for this situation is
obtained from \eqref{square rel} by replacing 
$Y_k$ with $Y_k\sum_{j=0}^k X_j$, and by rewriting
the resulting right-hand side by using \eqref{square rel}
together with some manipulation. Doing this by
hand already becomes quite tedious; so we use \texttt{Sigma}
to carry out this task automatically:

\begin{mma}\MLabel{MMA:mySum2}
\In mySum2=\sum_{k=0}^a \Bigg(
        \sum_{j=0}^k X[j]\Bigg)^2 Y[k];\\
\end{mma}

\begin{mma}\MLabel{MMA:CalkinXY2}
\In res2=SigmaReduce[mySum2,XList\to\{X,Y\},XWeight\to\{2,1\},\newline
 \hspace*{2cm} SimplifyByExt\to DepthNumberDegree,SimpleSumRepresentation\to True]\\
\Out -2 
\sum_{i=0}^a \Bigg(
        \sum_{j=0}^i X[j]
\Bigg)
\Bigg(
        \sum_{j=0}^i Y[j]\Bigg) X[i]
+2 
\sum_{i=0}^a \Bigg(
        \sum_{j=0}^i X[j]\Bigg) X[i] Y[i]\newline
\hspace*{1cm}+
\sum_{i=0}^a \Bigg(
        \sum_{j=0}^i Y[j]\Bigg) X[i]^2
+\Bigg(
        \sum_{i=0}^a X[i]\Bigg)^2 \Bigg(
        \sum_{i=0}^a Y[i]
\Bigg)
-
\sum_{i=0}^a X[i]^2 Y[i]\\
\end{mma}

\begin{psremark}\label{Remark:XYCalkin2}
If we execute \texttt{SigmaReduce} with the same options as described in Remark~\ref{Remark:XYCalkin1}, we would fail for this input sum: there is no alternative expression in terms of nested sums where the nesting depth is simpler -- even with the assumption that $X[k]$ is considered as more nested than $Y[k]$ \footnote{If a simpler expression exists, \texttt{Sigma} would find it with the same options as described in Remark~\ref{Remark:XYCalkin1}.}. However, inserting the extra option \MInText{SimplifyByExt\to DepthNumberDegree} one aims at a simplification where the degree of the most complicated sum $\sum_{j=0}^k X[j]$ in~\myIn{\ref{MMA:mySum2}} is minimized; in addition, extra sums with lower nesting depth will be used (exploiting the fact that $Y[j]$ is less nested than $X[j]$) whenever such a degree reduction can be performed.
This simplification strategy can be set up by combining the enhanced telescoping algorithms from~\cite[Section~5]{Schneider:07d} with~\cite{FastAlgorithm3} to make \texttt{Sigma} compute~\myOut{\ref{MMA:CalkinXY2}} as an
alternative presentation of
\begin{equation}\label{xy version power of 2 sum}
\sum_{k=0}^a Y_k\left(\sum_{j=0}^k X_j\right)^2.
\end{equation}

\end{psremark}

Specializing $Y_k=1$ and $X_j= \binom{n}{j}$
in this generic relation~\myOut{\ref{MMA:CalkinXY2}} gives
\begin{equation} \label{C2}
C_2(a,n)=(a+1)\left(\sum_{k=0}^a \binom{n}{k}\right)^2
- 2 \sum_{k=0}^a k \binom{n}{k} \sum_{j=0}^k \binom{n}{j}
+ \sum_{k=0}^a k \binom{n}{k}^2.
\end{equation}
The specialization $a=n$ is treated algorithmically in Subsection~\ref{Sec:Calkin2Simple} resulting in the presentation~\eqref{C2 Vn} for $C_2(n)$.

The paper is organized as follows. After introducing the basic notions and constructions for setting up summation problems in terms of generic sequences in Section~\ref{Sec:Generic}, in Section~\ref{Sec:BasicTactics} we explain the basic simplification machinery to reduce double sums to expressions in terms of single nested sums. In Section~\ref{Sec:DRAlg} we reformulate this simplification methodology in the setting of abstract difference rings, and in Section~\ref{Sec:SequenceSetting} we connect these ideas with the ring of sequences utilizing an advanced difference ring theory; further supporting tools and notions (like \rpisiSE-rings) can be found in Section~\ref{Sec:Appendix} of the Appendix. Putting everything together will enable us to show that the suggested simplification strategy forms a complete algorithm for inputs that are given in terms of indefinite nested sums defined over hypergeometric products, multibasic products and their mixed versions. In Section~\ref{Sec:Sigma} we give further details how this simplification engine is implemented in the package \texttt{Sigma} and elaborate various concrete examples. In Section~\ref{Sec:Conclusion} the paper concludes by giving some pointers to future research.

\section{Generic sequences and sums}\label{Sec:Generic}

We want to model sequences and sums generically.
To this end we introduce a set $X$ of indeterminates
indexed over $\mathbb{Z}$ together with the ring of
multivariate polynomials in these symbols over $\KK$ \footnote{$\KK$ is a field of characteristic $0$.},
\begin{equation}\label{Equ:KXDef}
X := \{ X_j\}_{j\in \mathbb{Z}} \hbox{\, and\ }
\KK_X:=\KK[X].
\end{equation}
It will be convenient to 
consider bilateral sequences $f: \ZZ\rightarrow \KK_X,
j\mapsto f(j)$. The set of bilateral sequences
is denoted by $\KK_X^\ZZ$. In the following we 
only speak about ``sequences''; whether a sequence
is bilateral or not will be always clear from the context.

\begin{psconvention}\label{Conv:Convention}
We fix $k$ as a ``generic'' symbol which in
this article we overload
with three different meanings which will be 
always clear from the context:
\begin{itemize}
\item As in Section \ref{intro}, $k$ can stand for an integer; i.e., $k\in \mathbb{Z}$.

\item It stands for the bilateral sequence 
$k: \ZZ\rightarrow \KK_X, j\mapsto j$.

\item More generally, $k$ stands for a generic variable, 
respectively index; i.e., for a sequence 
$P=(P(j))_{j\in \ZZ} \in \KK_X^\ZZ$
we alternatively write $P(k)$ ($=P$); see Example~\ref{ex:neq}.
\end{itemize}
\end{psconvention}

\noindent In particular,
the latter meaning arises in generic sequences
and sums defined in Definitions~\ref{def:generic seqs} 
and~\ref{def:generic sums}, respectively.

\begin{psdefinition}[generic sequences] \label{def:generic seqs} The symbol \emph{$X_k$ with generic index $k$} and 
its shifted versions $X_{k+l}$, $l\in \ZZ$, denote
bilateral sequences in $\KK_X^\ZZ$
defined as $X_{k+l}:\ZZ \rightarrow \KK_X, j\mapsto X_{j+l}$. The \emph{set of all such generic sequences}
is denoted by the symbol ``$\{X_k \}$''; i.e.,
$\{X_k \}:=\{ X_{k+l}\}_{l\in \ZZ}$. 
\end{psdefinition}

\noindent The ring $\KK_X[k,\{X_k\}]$ of polynomials 
in $k$ and in generic sequences from $\{X_k \}$
is a subring of
the ring of sequences $\KK_X^\ZZ$ with the usual
(component-wise) plus and times.

\begin{psexample}
$P(k)=k^2 X_0 X_{k-1} X_{k+1} - k X_{-3} X_k^2 
+X_3-2\in \KK_X[k,\{X_k\}]$ represents the sequence 
$(p(j))_{j\in \ZZ}$,
\[
P(k):\ZZ\rightarrow \KK_X, j\mapsto
p(j)=j^2 X_0 X_{j-1} X_{j+1} - j X_{-3} X_j^2 
+X_3-2.
\]
\end{psexample}

\begin{pslemma} \label{helper 1}
Let $P(k)\in \KK_X[k,\{X_k\}]$ be such that 
\[
P(j)=0 \mbox{ for all } j\geq \mu
\]
for some $\mu \in \NN$. Then $P(k)=0$, the zero sequence. 
\end{pslemma}

\begin{proof}
The statement is obvious if one views $P(k)$ as
a polynomial in $k$ over the integral domain
$\KK_X[\{X_k\}]$.~\qed
\end{proof}

\begin{psdefinition}[generic sums] \label{def:generic sums}
Given $P(k)\in \KK_X^\ZZ$,
for $a,b \in \ZZ$ the \emph{generic sum $\sum_{l=a}^{k+b} P(l)$}
denotes a sequence in $\KK_X^\ZZ$ defined as
\begin{equation}\label{Equ:SumEv}
\sum_{l=a}^{k+b} P(l): \ZZ\rightarrow \KK_X, j\mapsto
\begin{cases}
\sum_{l=a}^{j+b} P(l), &\mbox{if } a\leq j+b\\
0, & \mbox{otherwise}.
\end{cases}
\end{equation}
\end{psdefinition}

\begin{psexample} \label{ex:neq}
For any $P(k)\in \KK_X^\ZZ$ and
\[
(f_P(k))_{k\in \ZZ}:= \sum_{l=0}^k P(l)
 - \sum_{l=0}^{k-1} P(l)
\]
one has
\[
f_P(j)=\begin{cases}
P(j), &\mbox{if } j\geq 0\\
0, &\mbox{otherwise}.
\end{cases}
\]
In other words, in the context of generic 
sequences and sums,
\begin{equation}\label{neq}
\sum_{l=0}^k P(l) - \sum_{l=0}^{k-1} P(l)
\neq P(k).
\end{equation}
\end{psexample}

This leads us to introducing an equivalence relation
``$\equiv$'' such that in situations as in 
Example~\ref{ex:neq},
\begin{equation}\label{eq}
\left[ \sum_{l=0}^k P(l) \right]
- \left[ \sum_{l=0}^{k-1} P(l)\right]
\equiv [P(k)],
\end{equation}
where we write $[f]$ for the equivalence class
of a sequence $f\in \KK_X^\ZZ$.
\begin{psdefinition}
For $f=(f(j))_{j\in \ZZ}, g=(g(j))_{j\in \ZZ}
\in \KK_X^\ZZ$ define
\[
f\equiv g\, :\Leftrightarrow \, \exists \lambda\in \ZZ:
f(j)=g(j) \mbox{ for all }j\geq \lambda.
\]
\end{psdefinition}

\noindent Obviously this introduces an equivalence relation
on $\KK_X^\ZZ$. Equivalence classes are denoted by
$[f]$, the set of equivalence classes by 
$\mathrm{Seq}(\KK_X)$; i.e.,
\[
\mathrm{Seq}(\KK_X)=\{ [f]: f\in \KK_X^\ZZ\}.
\]

\noindent Clearly, $\mathrm{Seq}(\KK_X)$ forms a commutative
ring with $1$, which is defined by extending the usual (componentwise)
sequence operations plus and times in an obvious
way by $[f]+[g]:=[f+g]$ and $[f][g]:=[f g]$.

\noindent The shift operator 
\begin{equation}\label{Equ:ShiftDef}
S: \mathrm{Seq}(\KK_X) \rightarrow \mathrm{Seq}(\KK_X),
[f] \mapsto S[f]:=[S f]
\end{equation}
where $Sf = (f(j+1))_{j\in \ZZ}$ 
if $f = (f(j))_{j\in \ZZ}$,
is a ring automorphism, a property which is inherited
from the shift operator on sequences from $\KK_X^\ZZ$.
For $f(k)=(f(j))_{j\in \ZZ} \in \KK_X^\ZZ$ and $m\in \ZZ$ we often write $f(k+m)$ instead of 
$S^m f(k)=(f(j+m))_{j\in \ZZ}$.

\smallskip
\noindent\textit{Convention.} If things are clear from the context,
for equivalence classes from $\mathrm{Seq}(\KK_X)$ we
will simply write $f$ instead of $[f]$. Nevertheless,
we will continue to use ``$\equiv$'' to express 
equality between equivalence classes. For example,
instead of \eqref{eq} we write,
\begin{equation}\label{eq new}
\sum_{l=0}^k P(l) 
- \sum_{l=0}^{k-1} P(l)
\equiv P(k).
\end{equation}
In the same spirit, given $f(k)\in \KK_X^\ZZ$ 
and $m\in \ZZ$, we will write 
\[
f(k+m) \mbox{ instead of } [f(k+m)],
\]
provided that the meaning $f(k+m)\in \mathrm{Seq}(\KK_X)$
is clear from the context.

Summation methods often rely on coefficient comparison. To apply this technique one usually exploits algebraic
independence; for instance, 
equivalence classes $[f]$ of generic sums like $f=\sum_{l=0}^k X_l\in  \KK_X^\ZZ$ are algebraically
independent over $(\KK_X[k,\{X_k\}],\equiv)$.\footnote{
The quotient ring of $\KK_X[k,\{X_k\}]$ subject to
the equivalence relation $\equiv$; this ring is 
a subring of $\mathrm{Seq}(\KK_X)$.} Slightly more generally, we prove the following 

\begin{pslemma}\label{alg indep of sums}
Let $P(k)\in \KK_X[k]$. Then 
\[
\left[ \sum_{l=0}^{k} P(l) X_l \right]
\mbox{ is transcendental over }
(\KK_X[k,\{X_k\}],\equiv).
\]
\end{pslemma}

\begin{proof}
For $F(k):=\sum_{l=0}^{k} P(l) X_l\in \KK_X^{\ZZ}$ suppose that
\begin{equation}\label{alg indep 1}
0\equiv q_0(k) + q_1(k) F(k) + \dots + q_d(k) F(k)^d
\end{equation}
for polynomials $q_i(k)\in \KK_X[k,\{X_k\}]$ with
$q_d(k)\not\equiv 0$.\footnote{This means that $q_d(k)$
is not equivalent to the $0$-sequence 
$(\dots, 0,0,0,\dots)\in \KK_X^\ZZ$.} Let $d\geq 1$
be the minimal degree such that a relation like 
\eqref{alg indep 1} holds. Denoting the sequence
on the right side of ~\eqref{alg indep 1} by 
$(f(j))_{j\in \ZZ}$, we have that there is
a $k_0\in \ZZ$ such that
\[
f(j)=0 \mbox{ for all } j\geq k_0.
\]
Define
\[
l_0:=\mathrm{max}\{l\in \ZZ: X_l \mbox{ divides some
monomial of some }q_i(k)\},
\]
and set 
\[
j_0:=\mathrm{max}\{0, k_0, l_0+1 \}.
\]
Then 
\begin{align*}
0 & = \mbox{ coefficient of } X_{j_0}^d \mbox{ in } f(j_0) = q_d(j_0) P(j_0)^d,\\
0 & = \mbox{ coefficient of } X_{j_0}^d \mbox{ in } f(j_0+1) = q_d(j_0+1) P(j_0+1)^d,\\
& \mbox{etc.}
\end{align*}
Since $P(k)\in \KK_X[k]$ has at most finitely many
integer roots (if any), there is a $\mu\in \NN$ 
such that
\[
q_d(j)=0 \mbox{ for all } j\geq \mu.
\]
Consequently, $q_d(k)\equiv 0$, a contradiction to
$q_d(k)\not \equiv 0$.
Therefore  $d=0$, and the statement follows
from Lemma \ref{helper 1}.\qed
\end{proof}

\section{The basic simplification}\label{Sec:BasicTactics}

In the following, instead of considering sums
like \eqref{xy version power of 2 sum}, we will restrict to a slightly less general class of sums by setting $Y_j=1$ for all $j\geq0$, i.e., we will explore for $p=1,2$ the sums
\begin{equation}\label{Equ:CalkinSumX}
\sum_{j=0}^a \left(\sum_{l=0}^j X_l\right)^p
\end{equation}
involving the generic sequence $X_k$.
Obviously, for fixed $p$ this sum can be viewed as a sequence
$s(a)=(s(a))_{a\in \ZZ}\in \KK_X^\ZZ$.\footnote{Note
that $s(a)=0$ if $a<0$.} So, 
more precisely, we will investigate if and how 
sequences from $\KK_X^\ZZ$ given by such sum
expressions can be simplified in terms of 
``simpler'' generic sums.

\subsection{Simplifications by sum extensions}\label{Sec:Calkin1}
We start to look at the case $p=1$ 
of~\eqref{Equ:CalkinSumX}, respectively $C_1(a,n)$,
by considering the following problem.

\medskip

\noindent\textit{Given} a generic sum $F(k)=\sum_{l=0}^k X_l\in \KK_X^\ZZ$;\\
\textit{find} $G(k)\in \KK_X^\ZZ$,
``\textit{as simple as possible}'', such that 
\begin{equation}\label{Equ:Tele}
G(k+1)-G(k)\equiv F(k+1).
\end{equation}

\vspace*{-0.3cm}

\noindent Trivially, 

\vspace*{-0.5cm}

\begin{equation}\label{trivial sol}
G(k)=\sum_{j=0}^k F(j)\in \KK_X^\ZZ
\end{equation}
is always a solution to \eqref{Equ:Tele}. So the problem
splits into two parts: (a) to specify a
concrete meaning of ``as simple as possible'', and (b)
to compute solutions which meet this specification. 

\medskip

For part (a), for the given problem we start by
considering solutions of the form
\begin{equation}\label{Equ:AnsatzG}
G(k)=G_0(k) + G_1(k) F(k)
\end{equation}
with $G_j(k)\in \KK_X[k,\{X_k\}]$ to be determined,
the latter task being part (b) of the problem.

\medskip

In practice the specifications given to settle
part (a) of the problem
are motivated by the context of the problem,
but also driven by theory. For instance, here
Lemma~\ref{alg indep of sums} implies that there
is no solution $G(k)\in \KK_X[k,\{X_k\}]$
to the telescoping equation \eqref{Equ:Tele}.
In this sense\footnote{By difference ring theory (see Lemma~\ref{Lemma:DegBound} below) the exponent with which $F(k)$ can appear in $G(k)$ is at most $2$. As it turns out, exponent $1$ suffices here to obtain a solution of the desired form.}, the ansatz in \eqref{Equ:AnsatzG} is the
best possible we can achieve.

\medskip

To execute part (b) of the problem we proceed
by coefficient comparison. To this end, we 
substitute the ansatz \eqref{Equ:AnsatzG} into
\eqref{Equ:Tele} to obtain:
\begin{align}\label{Equ:Tele1}
\left( G_1(k+1)-G_1(k) \right)& F(k) +
G_0(k+1)-G_0(k) + G_1(k+1) X_{k+1} \nonumber\\
& \equiv F(k)+X_{k+1}.
\end{align}
Owing to Lemma \ref{alg indep of sums} we can do
coefficient comparison with respect to powers 
of $F(k)$ and obtain,
\[
G_1(k+1)-G_1(k)\equiv 1.
\]
It is straightforward to verify that 
\[
G_1(k)=k+d, \mbox{\,  with } d\in \KK_X \mbox{ arbitrary},
\]
describes all the solutions in
$\KK_X[k,\{X_k\}]= \KK_X[\{X_k\}][k]$.
To keep things simple we set $d=0$, and
substituting $G_1(k)=k$ into \eqref{Equ:Tele1} yields
\begin{equation}\label{Equ:Tele2}
 G_0(k+1)-G_0(k)\equiv - k X_{k+1}.
\end{equation}
Using a similar idea as used in the proof
of Lemma \ref{alg indep of sums} reveals
that \eqref{Equ:Tele2} admits no solution
$G_0(k)\in \KK_X[k, \{X_k\}]$. So we are led
to relax our specification 
of ``simple'' and---
in view of~\eqref{trivial sol}---
set $G_0$ to the trivial solution of \eqref{Equ:Tele2}; i.e., to the generic sum 
\[
G_0(k)= - \sum_{j=0}^k j X_j + F(k)\, \, 
\Big(\equiv - \sum_{j=0}^{k} (j-1) X_{j} \Big).
\]
Putting things together, 

\vspace*{-0.7cm}

\begin{align}\label{sol 1}
G(k) &= G_0(k) + G_1(k) F(k) = - \sum_{j=0}^k j X_j + (k+1) F(k) \in \KK_X^\ZZ
\end{align}

\vspace*{-0.4cm}

\noindent is a solution of \eqref{Equ:Tele}.

Finally, we convert \eqref{Equ:Tele} into the form of a summation
identity.  Passing from the generic sequence variable
$k$ to concrete integers $k\in \ZZ$, using~\eqref{sol 1} 
we can easily verify that for all $k\geq 0$,
\begin{align*}
G(k)-G(k-1)&=-k X_k+(k+1)F(k)-k F(k-1)\\
&= -k X_k + (k+1)(F(k-1)+X_k)-k F(k-1)\\
&= X_k +F(k-1) = F(k).\footnotemark
\end{align*}
\footnotetext{Note that $F(-1)=0$ by definition of
a generic sum.}
Summing this telescoping relation over $k$ from $0$ to $a\in \ZZ$, 
$a\geq 0$, produces\footnote{According to \eqref{sol 1}: 
$G(-1)=0$.}
\begin{align*}
\sum_{k=0}^a \sum_{j=0}^k X_j
&= \sum_{k=0}^a F(k) =G(a)-G(-1)=G(a)\\
&= - \sum_{j=0}^a j X_j + (a+1) F(a)
= - \sum_{j=0}^a j X_j + (a+1) \sum_{j=0}^a X_j.
\end{align*}

Finally, observe that the generic sequence $X_k$ can be replaced by any concrete sequence $(\bar{X}_k)_{k\geq0}$ with $\bar{X}_k\in\KK$ yielding the identity
\begin{align}\label{Equ:Calkin1Simple}
\sum_{k=0}^a \sum_{j=0}^k \bar{X}_j= - \sum_{j=0}^a j \bar{X}_j + (a+1) \sum_{j=0}^a \bar{X}_j.
\end{align}

\medskip

With \texttt{Sigma} this can be obtained automatically. Namely, the package allows one to activate the desired mechanism by entering the sum 

\begin{mma}
 \In mySum=\sum_{k=0}^a\sum_{j=0}^kX[j];\\
\end{mma}

\noindent and executing the function call

\begin{mma}\MLabel{MMA:SigmaReduceXList}
 \In SigmaReduce[mySum,XList\to\{X\},SimpleSumRepresentation\to True]\\
 
 \vspace*{-0.1cm}
 
 \Out (a+1)\,\sum_{i=0}^{a}X_{i}-\sum_{i=0}^{a}iX_{i}\\
 \end{mma}

\subsection{Simplifications by introducing constraints and sum extensions}\label{Sec:Calkin2Simple}
Next, in view of the sum
\[
\sum_{k=0}^a k \binom{n}{k} \sum_{j=0}^k \binom{n}{j},
\]
arising in the presentation \eqref{C2} for $C_2(a,n)$,
we look at the following problem.

\medskip

\noindent\textit{Given} a generic sum 
$F(k)=k\,X_k\sum_{j=0}^kX_j\in \KK_X^\ZZ$;\\
\textit{find} $G(k)\in \KK_X^\ZZ$,
\textit{as simple as possible}, such that 
\begin{equation}\label{Equ:Tele3}
G(k+1)-G(k)\equiv F(k+1).
\end{equation}

\medskip
This time we start by
considering solutions of the form
\begin{equation}\label{Equ:AnsatzG3}
G(k)=G_0(k) + G_1(k) S(k) + G_2(k) S(k)^2
\end{equation}
with $S(k):=\sum_{j=0}^k X_j$, and where 
we again try to find the coefficients $G_j(k)$ 
of polynomial form such that
$G_j(k)\in \KK_X[k,\{X_k\}]$.

\medskip
\noindent To this end, we again proceed
by coefficient comparison; i.e., we 
substitute the ansatz~\eqref{Equ:AnsatzG3} 
into~\eqref{Equ:Tele3} to obtain:
\begin{align}\label{Equ:Tele4}
\left( G_2(k+1)-G_2(k) \right) S(k)^2
& + \left( G_1(k+1)-G_1(k) + 2 G_2(k+1) X_{k+1} \right) S(k)\\
&+ G_0(k+1)-G_0(k)+ G_1(k+1) X_{k+1} + G_2(k+1) X_{k+1}^2
 \nonumber\\
& \equiv (k + 1) X_{k + 1} S(k)+(k + 1) X_{k + 1}^2.
\end{align}
\noindent Owing to Lemma \ref{alg indep of sums} we again can do
coefficient comparison. With respect to $S(k)^2$ we obtain,
\begin{equation}\label{Equ:Degree2}
G_2(k+1)-G_2(k)\equiv 0.
 \end{equation}
This has $G_2(k)=c$, $c\in \KK_X$ arbitrary,
as the general solution in
$\KK_X[k,\{X_k\}]= \KK_X[\{X_k\}][k]$.

\smallskip

\noindent Coefficient comparison
with respect to $S(k)$ in \eqref{Equ:Tele4} gives
\begin{equation}\label{Equ:Tele Y1}
 G_1(k+1)-G_1(k)\equiv (k+1-2 c) X_{k+1}.
\end{equation}
In order to proceed, we suppose that the generic sequence
$Y_k\in \KK_X^\ZZ$ is a solution to \eqref{Equ:Tele Y1} and set $G_1(k):=Y_k$.

\smallskip

\noindent Finally, coefficient comparison
with respect to $S(k)^0$ in \eqref{Equ:Tele4} gives
\begin{equation}\label{Equ:Tele5}
 G_0(k+1)-G_0(k)\equiv (k+1-c) X_{k+1}^2 - Y_{k+1} X_{k+1}.
\end{equation}
Similarly to the situation in equation \eqref{Equ:Tele2} we relax our specification 
of ``simple'' and set $G_0$ to the trivial solution of~\eqref{Equ:Tele5}; i.e., to the generic sum 
\begin{equation*}
G_0(k)= \sum_{j=0}^k (j-c) X_j^2  - \sum_{j=0}^k X_j Y_j.
\end{equation*}

\smallskip

\noindent Combining all these ingredients yields the solution
\begin{equation}\label{Equ:Calk2VarTeleSol}
G(k)=c\, \big(\sum_{j=0}^k X_j \big)^2+Y_k\,\sum_{j=0}^kX_j+\sum_{j=0}^k(-c X_{j}^2+j X_{j}^2-X_{j} Y_{j}) \in \KK_X^\ZZ,
\end{equation}
under the assumption that 
\begin{equation}\label{y constraint 1}
Y_k\in \KK_X^\ZZ \mbox{ and } c\in \KK_X 
\mbox{ are chosen so that \eqref{Equ:Tele Y1} holds.}
\end{equation}

Finally, as in Subsection \ref{Sec:Calkin1}
we convert \eqref{Equ:Tele3} into a summation
identity.  Passing from the generic sequence variable
$k$ to concrete integers $k\in \ZZ$, using 
\eqref{Equ:Calk2VarTeleSol} 
we can easily verify that telescoping yields
for all integers $a\geq 0$,
\begin{equation}\label{Equ:CalkinVarOneSum2}
\sum_{k=0}^a k\,X_k\sum_{j=0}^k X_j=c\, 
\big( \sum_{j=0}^a X_j \big)^2
-c \sum_{j=0}^a X_j^2
- \sum_{j=0}^a X_j Y_j
+ Y_a \sum_{j=0}^a X_j
+ \sum_{j=0}^a j X_j^2
\end{equation}
under the constraint that the sequence values
$Y_k\in \KK_X$ and $c\in\KK_X$ are chosen such
\begin{equation}\label{Equ:Tele Y1 sequence version}
 Y_{k+1}-Y_{k} = (k+1-2 c) X_{k+1} \mbox{ for all }
 k \geq 0.
\end{equation}

Using \texttt{Sigma} this solution strategy can be automatically applied to the sum

\begin{mma}
 \In mySum=\sum_{k=0}^ak\,X[k]\sum_{j=0}^kX[j];\\
\end{mma}

\noindent with the procedure call\footnote{By using the option \texttt{RefinedForwardShift$\to$False}, Sigma follows the calculation steps carried out above. Without this option a more complicated (but more efficient) strategy is used that produces a slight variation of the output.}

\begin{mma}\MLabel{MMA:SigmaReduceNotSimpleWithConstraints}
 \In \{closedForm,constraint\}=SigmaReduce[mySum,XList\to\{X\},ExtractConstraints\to\{Y\},\newline
 \hspace*{3cm}SimpleSumRepresentation\to False,RefinedForwardShift\to False]\\
 \Out \{ c\,(\sum_{i=0}^aX[i])^2+Y[a]\,\sum_{i=0}^aX[i]+\sum_{i=0}^a(-c X[{i}]^2+i X[{i}]^2-X[{i}] Y[{i}]),\newline
\{Y[{a+1}]-Y[{a}]=(1+a)X[{a+1}]-2\,c\,X[{a+1}]\}\}\\
\end{mma}

\smallskip

\noindent This yields the identity~\eqref{Equ:Calk2VarTeleSol} with the constraint~\eqref{Equ:Tele Y1 sequence version}.\\ 
To produce the output in exactly the same form
as in identity~\eqref{Equ:CalkinVarOneSum2}, one can use the option \texttt{SimpleSumRepresentation$\to$True} to the derived result:

\begin{mma}\MLabel{MMA:SigmaReduceSimpleWithConstraints}
 \In SigmaReduce[closedForm, a, XList\to\{X,Y\}, SimpleSumRepresentation\to True]\\
 \Out c \big(
        \sum_{i=0}^a X[i]\big)^2
-c 
\sum_{i=0}^a X[i]^2
-
\sum_{i=0}^a X[i] Y[i]
+\big(
        \sum_{i=0}^a X[i]\big) Y[a]
+
\sum_{i=0}^a i X[i]^2\\
\end{mma}

\medskip

\noindent Further details on the calculation steps in the setting of difference rings will be given in Subsection~\ref{Sec:SymbolicApproach}.

\bigskip

As a consequence, one can now fabricate specialized identities with the following strategy.
Choose a concrete sequence $\bar{X}_k\in\KK$ such that one finds a ``nice'' solution  
$\bar{Y}_k\in\KK$ and $c\in\KK$ for 
\begin{equation}\label{Equ:PTele}
\bar{Y}_{k+1}-\bar{Y}_{k}=(1+k)\bar{X}_{k+1}-c\,2\bar{X}_{k+1}.
\end{equation}
This will yield the specialized identity
\begin{equation}\label{Equ:CalkinVarOneSum2Ev}
\sum_{k=0}^a k\,\bar{X}_k\sum_{j=0}^k \bar{X}_j=c\, 
\big( \sum_{j=0}^a \bar{X}_j \big)^2
-c \sum_{j=0}^a \bar{X}_j^2
- \sum_{j=0}^a \bar{X}_j \bar{Y}_j
+ \bar{Y}_a \sum_{j=0}^a \bar{X}_j
+ \sum_{j=0}^a j \bar{X}_j^2.
\end{equation}

\begin{psexample}\label{Exp:CalkinVar2}
Taking $\bar{X}_k=\binom{n}{k}$ in \eqref{Equ:CalkinVarOneSum2Ev} leads
to solving
\begin{equation}\label{Equ:Tele Y1 sequence version a}
 \bar{Y}_{k+1}-\bar{Y}_{k} = (k+1-2 c)\binom{n}{k+1} \mbox{ for all }
 k \geq 0.
\end{equation}

\noindent which can be done by \texttt{Sigma} as follows:

\begin{mma}\MLabel{MMA:Para1}
\In ParameterizedTelescoping[\{(k+1)SigmaBinomial[n,k+1],-2 SigmaBinomial[n,k+1]\},k]\\
\Out \{\{1,\frac{n}{4}, -\frac{1}{2}(k+1) \binom{n}{k+1}\}\}\\
\end{mma}

\medskip

\noindent The output \myOut{\ref{MMA:Para1}} means that as a solution to \eqref{Equ:Tele Y1 sequence version a} we have
\[
\bar{Y}_k = -\frac{1}{2}(k+1) \binom{n}{k+1}
=-\frac{1}{2} \binom{n}{k} (n-k)
\mbox{ and } c=\frac{n}{4}.
\]

\medskip

\noindent\textit{Remark.} Alternatively, one can use the RISC package \texttt{fastZeil}~\cite{Paule:95} by
\begin{mma}
\In << RISC`fastZeil` \\
\Print \LoadP{
Fast Zeilberger Package version 3.61
written by Peter Paule, Markus Schorn, and Axel Riese
\copyright RISC-JKU}\\
\end{mma}
\vspace*{0.2cm}

\begin{mma}\MLabel{MMA:Gosper}
\In Gosper[ Binomial[n, k+1], k, 1]\\

\vspace*{-0.2cm}

\Out {(-2 - 2 k + n) Binomial[n, 1 + k] == 
  \Delta_k[(1 + k) Binomial[n, 1 + k]]}\\
\end{mma}

\noindent\myIn{\ref{MMA:Gosper}}~calls an extended version
of Gosper's algorithm. In the given example
the last entry ``1'' asks the procedure to compute
- in case it exists - a polynomial $p_1(n) k + p_0(n)$
of order 1 in $k$ such that the polynomial times
the summand $\binom{n}{k+1}$ telescopes. In \myOut{\ref{MMA:Gosper}}
this polynomial is determined to be $(-2) k + n-2$;
$(\Delta_k f)(k)=f(k+1)-f(k)$ is the forward difference
operator.

\medskip

\noindent This turns \eqref{Equ:CalkinVarOneSum2Ev} into
\begin{align}\label{Equ:CalkinVar2a}
\sum_{k=0}^a k\, \binom{n}{k}\sum_{j=0}^k
 \binom{n}{j}
 & =\frac{n}{4}\, 
\Big( \sum_{j=0}^a \binom{n}{j} \Big)^2
+\frac{n}{4}\, \sum_{j=0}^a \binom{n}{j}^2 \\
&\, \, \,  + \frac{1}{2}\sum_{j=0}^a j\, \binom{n}{j}^2
-\frac{n-a}{2}\binom{n}{a} 
\sum_{j=0}^a \binom{n}{j}. \nonumber
\end{align}
For $a=n$ we have, using $\sum_{j=0}^m \binom{a}{j}
\binom{b}{m-j}=\binom{a+b}{m}$ and $\binom{n}{j}=
\frac{n}{j}\binom{n-1}{j-1}
=\frac{n}{j}\binom{n-1}{n-j}$, 

\begin{equation*}
\sum_{k=0}^n k\, \binom{n}{k}\sum_{j=0}^k
 \binom{n}{j}
  =\frac{n}{4}\, 2^{2n} 
+\frac{n}{4}\,\binom{2n}{n} 
  + \frac{n}{2} \binom{2n-1}{n} 
  = n\, 4^{n-1} + n \binom{2n-1}{n}.
\end{equation*}

\noindent Finally, substituting \eqref{Equ:CalkinVar2a} into
equation \eqref{C2} yields,
\begin{equation}\label{C2 Va} 
C_2(a,n)
=\left(a+1-\frac{n}{2}\right)
\left(\sum_{j=0}^a \binom{n}{j}\right)^2
- \frac{n}{2} \sum_{j=0}^a \binom{n}{j}^2
+ (n-a) \binom{n}{a} \sum_{j=0}^a \binom{n}{j}.
\end{equation}
Similarly to before, for $a=n$ this simplifies to
\begin{equation}\label{C2 Vn} 
C_2(n)=C_2(n,n)
=\left(\frac{n}{2} +1\right) 2^{2n}
- \frac{n}{2}  \binom{2 n}{n}
=(n+2) 2^{2n-1}-n \binom{2n-1}{n}.
\end{equation}
\end{psexample}

\begin{psexample}\label{Exp:Calkin S1}
Taking $\bar{X}_k=H_k:=\sum_{i=1}^k\frac1i$ in \eqref{Equ:CalkinVarOneSum2Ev} leads
to solving
\begin{equation*}
 \bar{Y}_{k+1}-\bar{Y}_{k} = (k+1-2 c)H_{k+1} \mbox{ for all }
 k \geq 0.
\end{equation*}
The solution
\[
\bar{Y}_k=\frac{1}{4} \big(
        -k^2
        +2 k (k+1) H_k
        +k
        -5
\big) \mbox{ and } c=0
\]
turns \eqref{Equ:CalkinVarOneSum2Ev} into 
\begin{align*}
\sum_{k=0}^ak\,H_k\sum_{j=0}^kH_j=&\frac{1}{4} \big(
        -5
        +a
        -a^2
        +2 a (a+1) H_a
\big) 
\sum_{j=0}^a H_j+
\sum_{j=0}^a j H_j^2\\
&-
\sum_{j=0}^a \frac{1}{4} \big(
        -5
        +j
        -j^2
        +2 j (1+j) H_j
\big) H_j\\
\stackrel{\texttt{Sigma}}{=}&-\tfrac{(2 a+1)(
        5 a^2+5 a-6)}{18}H_a 
+\tfrac{a(
        20 a^2+3 a-59)}{108} 
+\tfrac{a (a+1) (a+2)}{3} H_a^2.
\end{align*}

\noindent The second equality is obtained by applying \texttt{SigmaReduce} to the specialized expression. Here the underlying difference ring theory~\cite{DR3} is utilized in order to return an expression in terms of sums which are algebraically independent among each other.
\end{psexample}

\begin{psexample}\label{Exp:Calkin with squares}
Taking $\bar{X}_k=\binom{n}{k}^2$ in \eqref{Equ:CalkinVarOneSum2Ev} leads
to solving 
\begin{equation*}
 \bar{Y}_{k+1}-\bar{Y}_{k} = (k+1-2 c)\binom{n}{k+1}^2 \mbox{ for all }
 k \geq 0.
\end{equation*}
The solution
\[
\bar{Y}_k=-\frac{(n-k)^2}{2 n}\binom{n}{k}^2 
\mbox{ and } c=\frac{n}{4}
\]
turns \eqref{Equ:CalkinVarOneSum2Ev} into 
\begin{align*}
\sum_{k=0}^ak\,\binom{n}{k}^2\sum_{j=0}^k\binom{n}{j}^2=&-\frac{\binom{n}{a}^2}{n} \frac{1}{2} (-a
+n
)^2 
\sum_{j=0}^a \binom{n}{j}^2
+\frac{1}{4} n \big(
        \sum_{j=0}^a \binom{n}{j}^2\big)^2\\
&-\frac{1}{4} n 
\sum_{j=0}^a \binom{n}{j}^4
+
\sum_{j=0}^a j \binom{n}{j}^4
-
\sum_{j=0}^a -\frac{\binom{n}{j}^4 (-j
+n
)^2}{2 n}
\\
\stackrel{\texttt{Sigma}}{=}&\frac{
        -a^2
        +2 a n
        -n^2}{2n}  
\binom{n}{a}^2\sum_{i=0}^a \binom{n}{i}^2\\
&+\frac{1}{2n} 
\sum_{i=0}^a i^2 \binom{n}{i}^4
+\frac{n}{4}\Big(
        \sum_{i=0}^a \binom{n}{i}^2\Big)^2
+\frac{n}{4} 
\sum_{i=0}^a \binom{n}{i}^4
\end{align*}
which holds for all $a,n\in\NN$ with $n\neq0$.
\end{psexample}



\section{A reformulation in abstract difference rings}\label{Sec:DRAlg}

In the following we plan to gain more insight into when the double sums under consideration can be simplified to single sums. So far, we showed that the double sum on the left-hand side of~\eqref{Equ:CalkinVarOneSum2Ev} in terms of a sequence $(\bar{X}_k)_{k\geq0}$ with $\bar{X}_k\in\KK$ can be simplified to the right-hand side of~\eqref{Equ:CalkinVarOneSum2Ev} in terms of single nested sums provided that for $c\in\KK$ and $\bar{Y}_k\in\KK$ the parameterized telescoping equation~\eqref{Equ:PTele} holds. In the following we will show that for certain classes of sequences $\bar{X}_k$ and $\bar{Y}_k$ the constraint~\eqref{Equ:PTele} is not only sufficient but also necessary; see Theorem~\ref{Thm:SeqStatement} below.
In order to accomplish this task, we will utilize new results of difference ring theory~\cite{DR1,DR2,DR3,DR4}; compare also~\cite{Singer:97}. To warm up, we first rephrase the constructions of the previous sections in the difference ring setting.

\begin{psdefinition}
A \emph{difference ring (resp.\ field)} $\dfield{\AA}{\sigma}$ is a ring (resp.\ field) $\AA$ equipped with a ring (resp.\ field) automorphism $\fct{\sigma}{\AA}{\AA}$.
\end{psdefinition}

\noindent In fact, in Section~\ref{Sec:Generic} we introduced the difference ring $\dfield{\mathrm{Seq}(\KK_X)}{\Shift}$ where $\mathrm{Seq}(\KK_X)$ is the ring of (equivalent) sequences equipped with the ring automorphism defined in~\eqref{Equ:ShiftDef}. In addition, we considered the subring $\AA_1:=(\KK_X[k,\{X_k\}],\equiv)$ of $\mathrm{Seq}(\KK_X)$. Since $\AA_1$ is closed under $S$, the restricted version of $S$ to $\AA_1$ forms a ring automorphism. In short, we obtain the difference ring $\dfield{\AA_1}{\Shift}$ which is a subdifference ring of $(\mathrm{Seq}(\KK_X),\Shift)$.

\begin{psdefinition}
A difference ring $\dfield{\AA'}{\sigma'}$ is called a \emph{subdifference ring} of $\dfield{\AA}{\sigma}$ if $\AA'$ is a subring of $\AA$ and $\sigma'(a)=\sigma(a)$ for all $a\in\AA'$. Conversely, $\dfield{\AA}{\sigma}$ is called a \emph{difference ring extension} of $\dfield{\AA'}{\sigma'}$. Since $\sigma'$ agrees with $\sigma$ on $\AA'$, we usually do not distinguish anymore between them.
\end{psdefinition}

\noindent Further, by Lemma~\ref{alg indep of sums}
the sequence $\sum_{l=0}^{k} X_l\in\mathrm{Seq}(\KK_X)$ is transcendental over $\AA_1$. Thus the smallest subring of $\mathrm{Seq}(\KK_X)$ that contains $\AA_1$ and $\sum_{l=0}^{k}  X_l$ forms a polynomial ring which we denote by
\begin{equation}\label{Equ:A2Def}
\AA_2:=\KK_X[k,\{X_k\}]\left[ \sum_{l=0}^{k}X_l \right].
\end{equation}
Then using the fact that
\begin{equation}\label{Equ:A2ShiftDef}
S\sum_{l=0}^{k}X_l\equiv\sum_{l=0}^{k+1}X_l \equiv
\sum_{l=0}^{k}X_l+X_{l+1}
\end{equation}
holds with $X_{l+1}\in\KK_X[k,\{X_k\}]$ 
it follows that $\AA_2$ is closed under $S$ and thus $\dfield{\AA_2}{\Shift}$ is a subdifference ring of $\dfield{\mathrm{Seq}(\KK_X)}{\Shift}$. Summarizing, we obtain the following chain of difference ring extensions:
$$\dfield{\KK_X}{\Shift}\leq\dfield{\AA_1}{\Shift}\leq\dfield{\AA_2}{\Shift}\leq\dfield{\mathrm{Seq}(\KK_X)}{\Shift}$$
where $\dfield{\KK_X}{\Shift}$ is the trivial difference ring with $\Shift(f)\equiv f$ for all $f\in\KK_X$, i.e., the elements in $\KK_X$ are precisely the constant sequences.

In the light of these constructions, we can reformulate the problem in Subsection~\ref{Sec:Calkin2Simple} within the difference ring $\dfield{\AA_2}{\Shift}$ as follows: Given the sequence
$F(k)=k\,X_k\sum_{j=0}^kX_j\in\AA_2$, find a sequence $G(k)\in\AA_2$ or in a suitable subring of $\mathrm{Seq}(\KK_X)$ such that
$$G(k+1)-G(k)\equiv F(k).$$
Here we found out that we can choose~\eqref{Equ:Calk2VarTeleSol} with $Y_k\in\mathrm{Seq}(\KK_X)$ and $c\in\KK_X$ which satisfies the constraint~\eqref{y constraint 1}.
Thus specializing $X_k$ to concrete sequences $(\bar{X}_k)_{k\geq0}$ with $\bar{X}_k\in\KK$ such that there is a nice sequence $(\bar{Y}_k)_{k\geq0}$ with $\bar{Y}_k\in\KK$ that satisfies property~\eqref{Equ:PTele} for some $c\in\KK$ will lead to the simplification~\eqref{Equ:CalkinVarOneSum2Ev}. 

In the following we denote by $\seqK$ the subset of all sequences of $\mathrm{Seq}(\KK_X)$ whose entries are from $\KK$. Then it follows that $\seqK$ is a subring of $\mathrm{Seq}(\KK_X)$ and that $\Shift:\mathrm{Seq}(\KK_X)\to\mathrm{Seq}(\KK_X)$ restricted to $\seqK$ forms a ring automorphism. Thus $(\seqK,\Shift)$ forms a subdifference ring of $(\mathrm{Seq}(\KK_X),\Shift)$. Sometimes  $(\mathrm{Seq}(\KK_X),\Shift)$ is also called the \emph{difference ring of sequences}.

\begin{psremark}
Usually, the difference ring $(\seqK,\Shift)$ is defined by starting with the commutative ring $\KK^{\NN}$ with $1$ and defining the equivalence relation 
\[
f\equiv g\, :\Leftrightarrow \, \exists \lambda\in \NN:
f(j)=g(j) \mbox{ for all }j\geq \lambda
\] 
for $f=(f(j))_{j\geq0}, g=(g(j))_{j\geq0}\in \KK^{\NN}$; compare~\cite{AequalB}. It is easily seen that the set of equivalence classes $[f]$ with $f\in\KK^{\NN}$ forms a commutative ring with $1$ which is isomorphic to $\seqK$. In a nutshell, we can either choose $(a_n)_{n\in\NN}$ or $(a_n)_{n\in\ZZ}$ in order to describe the equivalence classes of $\seqK$. 
\end{psremark}

Subsequently, we will pursue a more general and ambitious goal. Namely, we will show that our new method produces constraints given in terms of parameterized telescoping equations that provide not only sufficient but also necessary conditions in order to simplify a nested sum in terms of generic sequences to an expression in terms of single nested sums over the given summand objects. 
In order to derive this extra insight, we will consider not an arbitrary specialization of $X_k,Y_k$ to general sequences $(\bar{X}_k)_{k\geq0},(\bar{Y}_k)_{k\geq0}\in\seqK$ but only to those sequences that can be generated by expressions in terms of indefinite nested sums defined over products. Typical examples are, e.g., the left- and right-hand sides of~\eqref{Equ:CalkinVar2a}, 
and~\eqref{C2 Va}; for a more precise definition we refer to Definition~\ref{Equ:DefNestedSums} below. With this restriction, we will then utilize Schneider's newly established difference ring results~\cite{DR1,DR2,DR3,DR4} to show that~\eqref{Equ:CalkinVarOneSum2Ev} is the only possible simplification of a double sum in terms of single sums.



In Schneider's difference ring approach sequences are represented by elements from a ring $\AA$ which is given either by certain rational function field extensions, polynomial ring extensions or by polynomial ring extensions factored out by certain ideals. In addition, a so-called evaluation function $\fct{\ev}{\AA\times\NN}{\KK}$ accompanies this ring construction that links the generators (variables) of the ring to the sequence interpretation. We will not give a full account on all the construction aspects~\cite{DR1,DR3}, but will emphasize  only the key steps that are relevant for our considerations below. Further details can be found in the Appendix~\ref{Sec:Appendix} below.

\begin{psexample}\label{Exp:Q(k)}
Consider the rational function field $\AA=\KK(k)$ in the variable $k$. Then we define the \emph{evaluation function} $\fct{\ev}{\AA\times\NN}{\KK}$ by 
\begin{equation}\label{Equ:EvalRat}
\ev(\tfrac{p}{q},i)=\begin{cases}
0&\text{if }q(i)=0\\
\frac{p(i)}{q(i)}&\text{if }q(i)\neq0;
\end{cases}
\end{equation}
where $p,q\in\KK[k]$ are polynomials with $q\neq0$; here $p(i),q(i)$ are the usual evaluations of polynomials at $i\in\NN$. Note that here we introduce yet another meaning of $k$, different from those introduced in Convention~\ref{Conv:Convention}: $k$ is an algebraic variable (indeterminate) that produces the rational function field $\KK(k)$. E.g., $f=1+k+k^2$ in this context is considered as a polynomial in 
the variable $k$ with integer coefficients and $s=(\ev(f,i))_{i\geq0}\in\seqK$ provides us with the corresponding sequence interpretation. With our earlier notations from
Convention~\ref{Conv:Convention} we could simply write $P(k)=1+k+k^2$ to abbreviate the same sequence $s$. 
\end{psexample}

Besides such a ring $\AA$, also a ring automorphism $\fct{\sigma}{\AA}{\AA}$ is introduced which scopes the shift behavior accordingly: for any $x\in\AA$ we will take care that 
\begin{equation}\label{Equ:EvHomomShift}
(\ev(\sigma(x),i))_{i\geq0}\equiv(\ev(x,i+1))_{i\geq0}=(\ev(x,i))_{i\geq1}
\end{equation}
holds. In addition, the construction is carried out so that the \emph{set of constants}\footnote{Note that $\const{\AA}{\sigma}$ in general is a subring of $\AA$.} 
$$\const{\AA}{\sigma}=\{c\in\AA|\,\sigma(c)=c\}$$
of the difference ring $\dfield{\AA}{\sigma}$ equals precisely the field $\KK$ in which the sequences are evaluated. All these properties hold, for instance, for the ground field $\AA=\KK(k)$ given in Example~\ref{Exp:Q(k)}.

\begin{psexample}\label{Exp:ModelXk}
Consider for instance the sequence $(\bar{X}_{i})_{i\geq0}$ with $\bar{X}_{0}=0$ and $\bar{X}_{i}=\frac{1}{i}$ for $i\geq1$.
Then we can choose the rational function $x:=\frac1k\in\AA$. In particular, we get~\eqref{Equ:EvHomomShift}. Further, we have $\KK=\const{\KK(k)}{\sigma}$.
\end{psexample}

In the following we will reconsider the calculation steps of Section~\ref{Sec:BasicTactics} within such abstract difference rings. In this context we will consider $X_k$ not as a generic sequence, but as a sequence $(\bar{X}_i)_{i\geq0}\in\seqK$ which can be modeled by an element $x\in\AA$ of a given difference ring
$\dfield{\AA}{\sigma}$ with $\KK=\const{\AA}{\sigma}$. 

\begin{psdefinition}
Let $\dfield{\AA}{\sigma}$ be a difference ring  with constant field $\KK$ and equipped with an \emph{evaluation function} $\ev$ satisfying~\eqref{Equ:EvHomomShift}. We say that a \emph{sequence $\bar{X}_k\in\KK$ is modeled by $x\in\AA$} if $\bar{X}_k=\ev(x,k)$ for all $k$ from a certain point on.
\end{psdefinition}

\noindent In particular, $\bar{X}_{k+i}$ with $i\in\ZZ$ is then modeled by $\sigma^i(x)\in\AA$. 
What we understand by ``modeled by'' has been illustrated also in the Example~\ref{Exp:ModelXk}.

\begin{psremark}
Note that the generic aspect is moved from a generic sequence $X_k$ to a ``generic'' difference ring $\dfield{\AA}{\sigma}$ and choosing an $x\in\AA$ from this ring $\AA$.
This change of paradigm will be very useful in Section~\ref{Sec:SequenceSetting} in order to show that the found simplifications are optimal in the sequence world. 
\end{psremark}

Next we explain how to adjoin the formal sum\footnote{Note that $\KK\subseteq\KK_X$ and thus the evaluation of a sum has been defined already in~\eqref{Equ:SumEv}.} 
\begin{equation}\label{Equ:InnerSum}
\sum_{i=0}^k\bar{X}_i
\end{equation}
to such an arbitrary ring $\AA$ with the shift behavior 
\begin{equation}\label{Equ:XSumShift}
\sum_{i=0}^{k+1}\bar{X}_i\equiv\sum_{i=0}^k\bar{X}_i+\bar{X}_{k+1}.
\end{equation}
To this end, we introduce a new variable $s$ being transcendental over $\AA$ and consider the polynomial ring $\AA[s]$. More precisely, using the fixed element $x\in\AA$, we define
\begin{equation}\label{Equ:Ev(s)}
\ev(s,i):=\sum_{j=1}^i \ev(x,j)=\sum_{j=1}^i\bar{X}_j
\end{equation}
in order to give $s$ the sequence meaning of our sum~\eqref{Equ:InnerSum}. More precisely, we extend this definition of $s$ to $\AA[s]$ by 
\begin{equation}\label{Equ:Ev(A[s])}
\ev(\sum_{l=0}^d f_l\,s^l,i)=\sum_{l=0}^d\ev(f_l,i)\ev(s,i)^l
\end{equation}
for any polynomial $\sum_{l=0}^d f_l\,s^l\in\AA[s]$ with $f_l\in\AA$.\\
Finally, we extend also the automorphism $\fct{\sigma}{\AA}{\AA}$ 
to  $\fct{\sigma'}{\AA[s]}{\AA[s]}$ with $\sigma'(h)=\sigma(h)$ for all $h\in\AA$ and 
\begin{equation}\label{Equ:sSumShift}
\sigma'(s)=s+\sigma(x).
\end{equation}
Note that to define the shift operator, we again used the fixed element $x\in\AA$.
More precisely, there is exactly one such automorphism where for $f=\sum_{l=0}^df_l\,s^l$ we obtain the map
$$\sigma'(f)=\sum_{l=0}^d\sigma(f_l)(s+\sigma(x))^l;$$
since $\sigma$ and $\sigma'$ agree on $\AA$, we do not distinguish them anymore.
In particular, by our construction it follows that 
$$(\ev(\sigma(f),i))_{i\geq0}\equiv(\ev(f,i+1))_{i\geq0}=(\ev(f,i))_{i\geq1}$$
for all $f\in\AA[s]$.

Summarizing, we constructed a difference ring extension $\dfield{\AA[s]}{\sigma}$ of $\dfield{\AA}{\sigma}$  where $s$ models the sum~\eqref{Equ:InnerSum}: $\ev$ provides the sequence representation and $\sigma$ describes the corresponding shift behavior. 

\noindent Note that this abstract construction can be turned to concrete applications.

\begin{psexample}\label{Exp:K(k)[s]}
We specialize $\dfield{\AA}{\sigma}$ to $\AA=\KK(k)$ and $\sigma(k)=k+1$. Starting with this ring, we want to model the harmonic numbers $H_k=\sum_{i=1}^k\bar{X}_i$ with $\bar{X}_i=\frac1i$. 
Thus we set $x:=\frac1k$ and follow the above construction, i.e., we take the difference ring extension $\dfield{\AA[s]}{\sigma}$ of $\dfield{\AA}{\sigma}$ with $s$ being transcendental over $\AA$ and with $\sigma(s)=s+\beta$ where $\beta:=\sigma(x)=\frac1{k+1}$. Further, we extend $\ev$ from $\AA$ to $\AA[s]$ by~\eqref{Equ:Ev(s)} and~\eqref{Equ:Ev(A[s])}. For $f=k\,s$ this yields, e.g., $\ev(f,i)=i\,H_i$ for $i\geq0$. Moreover, we obtain $\ev(\sigma(f),i)=\ev((i+1)H_{i+1},i)=\ev(i\,H_i,i+1)$ for all $i\geq0$. In a nutshell, we have rephrased the sequence of harmonic numbers $H_k$ by $s$ in $\AA[s]$ where $\ev$ provides the sequence representation and $\sigma$ describes the corresponding shift behavior. 
\end{psexample}

We emphasize that this elementary construction is still too naive for our subsequent considerations. Namely, a key feature will be that  
\begin{equation}\label{Equ:ConstProp}
\const{\AA[s]}{\sigma}=\const{\AA}{\sigma}
\end{equation}
holds. Together with our earlier assumption that $\const{\AA}{\sigma}=\KK$ holds, this will imply that in $\dfield{\AA[s]}{\sigma}$ the set of constants is precisely $\KK$. We install this special construction in the form of a definition.

\begin{psdefinition}\label{Def:SigmaExt}
Let $\dfield{\AA[s]}{\sigma}$ be a difference ring extension of $\dfield{\AA}{\sigma}$ with $s$ being transcendental over $\AA$ and $\sigma(s)=s+\beta$ for some $\beta\in\AA$. Then this extension is called a \emph{\sigmaSE-extension} if~\eqref{Equ:ConstProp} holds.
\end{psdefinition}

\noindent In the following we will rely heavily on the following result~\cite[Thm.~2.12]{DR1}; for the field version see~\cite{Karr:81}.

\begin{pstheorem}\label{Thm:SigmaExt}
Let $\dfield{\AA}{\sigma}$ be a difference ring with constant field $\KK$ and let $\dfield{\AA[s]}{\sigma}$ be a difference ring extension of $\dfield{\AA}{\sigma}$ with $s$ being transcendental over $\AA$ and with $\sigma(s)=s+\beta$ where $\beta\in\AA$. Then this is a \sigmaSE-extension (i.e., $\const{\AA[s]}{\sigma}=\const{\AA}{\sigma}$) iff there is no $g\in\AA$ with $\sigma(g)=g+\beta$. 
\end{pstheorem}

\begin{psremark}
Consider the difference ring extension $\dfield{\AA_2}{\Shift}$ of $\dfield{\AA_1}{\Shift}$ with~\eqref{Equ:A2Def} and~\eqref{Equ:A2ShiftDef}. 
By Lemma~\ref{alg indep of sums} $\AA_2$ is a polynomial ring over the coefficient domain $\AA_1$. One can show that $\const{\AA_2}{\Shift}=\const{\AA_1}{\Shift}=\KK_X$ which implies that $\dfield{\AA_2}{\Shift}$ is a \sigmaSE-extension of $\dfield{\AA_1}{\Shift}$.
By Theorem~\ref{Thm:SigmaExt}\footnote{In the theorem we require that the set of constants form a field. However, if $\const{\AA[s]}{\sigma}=\const{\AA}{\sigma}$, to prove the non-existence of a telescoping solution one does not need to assume that $\const{\AA}{\sigma}$ is a field.} this implies that the generic sum $\sum_{i=0}^kX_k$ cannot be simplified via telescoping in the difference ring $\dfield{\AA_1}{\Shift}$. However, specializing $X_k$ to a particular sequence $(\bar{X}_k)_{k\geq0}$, the situation might be different. 
\end{psremark}

Let us turn back to our generic construction:  we are given an arbitrary difference ring $\dfield{\AA}{\sigma}$ in which we choose $x\in\AA$ which models the desired sequence $\bar{X}_k$.
Suppose that there exists\footnote{In \texttt{Sigma} the existence can be decided constructively by efficient telescoping algorithms~\cite{FastAlgorithm2,FastAlgorithm3} provided that $\dfield{\AA}{\sigma}$ is a simple \rpisiSE-ring; see Appendix~\ref{Sec:Appendix}.} a $g\in\AA$ such that $\sigma(g)=g+\sigma(x)$ holds. 
In this case one can model the sum~\eqref{Equ:InnerSum} having the shift-behavior as in~\eqref{Equ:XSumShift} 
by $g$ with $\sigma(g)=g+\beta$. In other words, the double sum on the left-hand side of~\eqref{Equ:Calkin1Simple}
turns into a single sum in $\dfield{\AA}{\sigma}$. In the following we will ignore this degenerated case and assume that such a $g$ does not exist.

More precisely, we suppose that we are given a difference ring $\dfield{\AA}{\sigma}$ with constant field $\KK$ with the following properties:

\vspace*{-0.2cm}

\begin{enumerate}
 \item $\const{\AA}{\sigma}=\KK$;
\item there is a $k\in\AA$ with $\sigma(k)=k+1$;
 \item the sequence $\bar{X}_k\in\KK$ for $k\geq0$ can be modeled by an $x\in\AA$;
 \item there is no $g\in\AA$ with $\sigma(g)=g+\sigma(x)$, i.e., we cannot represent the sum~\eqref{Equ:InnerSum} in $\dfield{\AA}{\sigma}$.  
\end{enumerate} 

\vspace*{-0.2cm}
 
\noindent The third assumption together with Theorem~\ref{Thm:SigmaExt} implies that one can construct the \sigmaSE-extension $\dfield{\AA[s]}{\sigma}$ of $\dfield{\AA}{\sigma}$ with $\sigma(s)=s+\sigma(x)$. This means that $\AA[s]$ is a polynomial ring and $\const{\AA[s]}{\sigma}=\KK$.

\begin{psexample}\label{Exp:K(k)[s]Sigma}
Consider our concrete difference ring extension $\dfield{\AA[s]}{\sigma}$ of $\dfield{\AA}{\sigma}$ from Ex.~\ref{Exp:ModelXk} with $\AA=\KK(k)$ and $\sigma(s)=s+\beta$ with $\beta=\frac{1}{k+1}$. Using \texttt{Sigma} (or, e.g., Abramov's or Gosper's algorithms~\cite{Abramov:71,Gosper:78,Paule:95}), one can verify that there is no $g\in\KK(k)$ with $\sigma(g)=g+\beta$. Hence by Theorem~\ref{Thm:SigmaExt} our extension is a \sigmaSE-extension.
\end{psexample}

Within such a difference ring setting the telescoping problem in Subsection~\ref{Sec:Calkin2Simple} can be rephrased as follows.

\medskip

\noindent\textit{Given} $\dfield{\AA[s]}{\sigma}$ with the properties (1)--(4) from above and $f=k\,x\,s\in\AA[s]$.\\
\textit{Find} a $g\in\AA[s]$ such that
\begin{equation}\label{Equ:TeleDF}
\sigma(g)-g=\sigma(f)
\end{equation}
holds (note: $\sigma(f)=(k+1)\sigma(x)(s+\sigma(x))$).

\medskip

\noindent Now we repeat the calculation steps of Subsection~\ref{Sec:Calkin2Simple} within this (more abstract) difference ring exploiting the following extra insight~\cite[Lemma~7.2]{DR1}. 

\begin{pslemma}\label{Lemma:DegBound}
Let $\dfield{\AA[s]}{\sigma}$ be a \sigmaSE-extension of $\dfield{\AA}{\sigma}$ and $f,g\in\AA[s]$ with $\sigma(g)-g=f$. Then $\deg(g)\leq\deg(f)+1$. 
\end{pslemma}

\noindent Thus any solution $g\in\AA[s]$ of~\eqref{Equ:TeleDF} must have the form
$$g=g_0+g_1\,s+g_2\,s^2;$$
compare~\eqref{Equ:AnsatzG}. Plugging $g$ into~\eqref{Equ:TeleDF} we get
$$\sigma(g_2)(s+\sigma(x))^2+\sigma(g_1\,s+g_0)-\big[g_2\,s^2+g_1\,s+g_0\big]=(k+1)\,\sigma(x)(s+\sigma(x)).$$
The polynomials on the left- and right-hand sides agree if they agree coefficient-wise. Thus 
comparing coefficients with respect to $s^2$, it follows that $\sigma(g_2)=g_2$  which implies that $g_2\in\KK$. Thus we take an undetermined parameter $c\in\KK$ and set $g_2:=c$. Using this information we get
\begin{multline}\label{Equ:Deg1DR}
\big[\sigma(g_1)(s+\sigma(x))+\sigma(g_0)\big]-\big[g_1\,s+g_0\big]\\
=(k+1)\sigma(x)(s+\sigma(x))+c\big[-\sigma(x)^2-2\sigma(x)\,s\big].
\end{multline}
Again by coefficient comparison with respect to $s$ we obtain the constraint
\begin{equation}\label{Equ:ConstraintDeg1DR}
\sigma(g_1)-g_1=(1+k-2c)\sigma(x);
\end{equation}
compare with~\eqref{Equ:Tele Y1}.
Now suppose we find a $c\in\KK$ and a $y\in\AA$ such that 
\begin{equation}\label{Equ:ConstraintFory}
\sigma(y)-y=(1+k)\sigma(x)-2\,c\sigma(x)
\end{equation}
holds. Consequently, we get the general solution $g_1=y+d$ of~\eqref{Equ:ConstraintDeg1DR} 
for some undetermined constant $d\in\KK$. Plugging the solution into~\eqref{Equ:Deg1DR} yields
\begin{equation}\label{Deg0DR}
\sigma(g_0)-g_0=(k+1-c)\sigma(x)^2-\sigma(x) \sigma(y)-d\,\sigma(x);
\end{equation}
this is equivalent to~\eqref{Equ:Tele5} when $d=0$. At this point two scenarios may happen.\\
\smallskip
\noindent\textit{Case 1.} We find a $g_0\in\AA$ and $d\in\KK$ such that~\eqref{Deg0DR} holds. Then combining the derived sub-results provides the solution
\begin{equation}\label{Equ:gCase1}
g=c\,s^2+(y+d)\,s+g_0.
\end{equation}

\smallskip

\noindent\textit{Case 2.} We do not find a $g_0\in\AA$ and $d\in\KK$ such that~\eqref{Deg0DR} holds. Then we can construct the polynomial ring $\AA[s][t]$ and extend the automorphism $\sigma$ from $\AA[s]$ to $\AA[s][t]$ subject to the relation 
\begin{equation}\label{equ:DefinetShift}
\sigma(t)=t+\Big(\sigma(x)^2-c \sigma(x)^2+k \sigma(x)^2-\sigma(x) \sigma(y)\Big).
\end{equation}
By Theorem~\ref{Thm:SigmaExt} it follows that this extension is a \sigmaSE-extension. Namely, we have
$\const{\AA[s][t]}{\sigma}=\KK$. This, in particular, implies the solution $g_0=t$ and $d=0$ for~\eqref{Deg0DR}.  
Finally, in this case, combining the obtained representations of the coefficients produces the solution
\begin{equation}\label{Equ:gCase2}
g=c\,s^2+y\,s+t
\end{equation}
within the difference ring $\dfield{\AA[s][t]}{\sigma}$ where $c\in\KK$ and $y$ are a solution of~\eqref{Equ:ConstraintFory}; compare with~\eqref{Equ:Calk2VarTeleSol}.

\smallskip

\noindent The previous considerations can be summarized as follows.

\begin{pstheorem}\label{Thm:DRStatement}
Let $\dfield{\AA}{\sigma}$ be a difference ring with constant field $\KK$ and with $k\in\AA$ where $\sigma(k)=k+1$. 
Let $\dfield{\AA[s]}{\sigma}$ be a \sigmaSE-extension of $\dfield{\AA}{\sigma}$ with $\sigma(s)=s+\sigma(x)$ for some $x\in\AA$. 
Then the following holds.
\begin{myEnumerate}
\item There is a $g\in\AA[s]$ with $\sigma(g)-g=\sigma(k\,x\,s)$ iff the following two statements hold:

\vspace*{-0.2cm}

\begin{enumerate}
 \item[(a)] there is a $y\in\AA$ and $c\in\KK$ with~\eqref{Equ:ConstraintFory},
 \item[(b)] and there is a $g_0\in\AA$ and $d\in\KK$ with~\eqref{Deg0DR} (where $c$ is the one from part (a)).
\end{enumerate}

\vspace*{-0.1cm}

If (a) and (b) hold, we get the solution $g$ as given in~\eqref{Equ:gCase1}.

\vspace*{0.1cm}

\item There is a \sigmaSE-extension $\dfield{\AA[s][t]}{\sigma}$ of $\dfield{\AA[s]}{\sigma}$ with $\sigma(t)-t\in\AA$ together with a $g\in\AA[s][t]\setminus\AA[s]$ with $\sigma(g)-g=\sigma(k\,x\,s)$ iff the following two statements hold:

\vspace*{-0.2cm}

\begin{enumerate}
\item[(a)] there is a $y\in\AA$ and $c\in\KK$ with~\eqref{Equ:ConstraintFory},
\item[(b)] there is no $g_0\in\AA$ and $d\in\KK$ with~\eqref{Deg0DR} (where $c$ is the one from part (a)).
\end{enumerate}
\smallskip

\vspace*{-0.2cm}

If (a) and (b) hold, we get the solution $g$ as given in~\eqref{Equ:gCase2} with~\eqref{equ:DefinetShift}.
\end{myEnumerate}
\end{pstheorem}

Part~2 of the theorem describes the situation where one can adjoin a \sigmaSE-extension with the generator $t$ in order to gain a parameterized telescoping solution for~\eqref{Deg0DR}. Using the following extra insight from difference ring theory, we can generalize this situation if one allows a tower of single nested \sigmaSE-extensions.

\begin{pstheorem}\label{Thm:DRStatementGeneral}
Let $\dfield{\AA}{\sigma}$ be a difference ring with constant field $\KK$ and with $k\in\AA$ where $\sigma(k)=k+1$. 
Let $\dfield{\AA[s]}{\sigma}$ be a \sigmaSE-extension of $\dfield{\AA}{\sigma}$ such that $\sigma(s)=s+\sigma(x)$ for some $x\in\AA$. 
Then there is a tower of \sigmaSE-extensions $\dfield{\AA[s][t_1]\dots[t_e]}{\sigma}$ of $\dfield{\AA[s]}{\sigma}$ with $\sigma(t_i)-t_i\in\AA$ for $1\leq i\leq e$ together with a $g\in\AA[s][t_1,\dots,t_e]\setminus\AA[s]$ with $\sigma(g)-g=\sigma(k\,x\,s)$ iff the following two statements hold:
\begin{enumerate}
\item[(a)] there is a $y\in\AA$ and $c\in\KK$ with~\eqref{Equ:ConstraintFory},
\item[(b)] there is no $g_0$ and $d\in\KK$ with~\eqref{Deg0DR} (where $c$ is the one from part (a)).
\end{enumerate}
\smallskip
If (a) and (b) hold, we obtain the solution $g$ as given in~\eqref{Equ:gCase2} with~\eqref{equ:DefinetShift} (i.e., $e:=1$ and $t_1:=t$).
\end{pstheorem}
\begin{proof}
If statements (a) and~(b) hold, we can take ~\eqref{equ:DefinetShift} and get the solution $g$ as given in~\eqref{Equ:gCase2}. What remains to show is the other direction. Suppose that there is a tower of \sigmaSE-extensions $\dfield{\AA[s][t_1]\dots[t_e]}{\sigma}$ of $\dfield{\AA}{\sigma}$ with $\beta_i=\sigma(t_i)-t_i\in\AA$ for $1\leq i\leq e$. Assume further that there is a $g\in\AA[s][t_1,\dots,t_e]\setminus\AA[s]$ with $\sigma(g)-g=\sigma(k\,x\,s)$. By~\cite[Prop.~1]{AS:15} it follows that 
\begin{equation}\label{Equ:gShape}
g=g'+\kappa_1\,t_1+\dots+\kappa_e\,t_e
\end{equation}
for some $g'\in\AA[s]$ and $(\kappa_1,\dots,\kappa_e)\in\KK^e\setminus\{(0,\dots,0)\}$. Take the polynomial ring $\AA[s][t]$ and extend $\sigma$ from $\AA[s]$ to $\AA[s][t]$ subject to the relation $\sigma(t)=t+h$ with $h:=\kappa_1\,\beta_1+\dots+\kappa_e\,\beta_e$. By construction we have that 
\begin{equation}\label{Equ:AgainSolInT}
\sigma(g'+t)-(g'+t)=\sigma(g')-g'+\kappa_1\,\beta_1+\dots+\kappa_e\,\beta_e=\sigma(g)-g=\sigma(k\,x\,s).
\end{equation}
Now suppose that $\dfield{\AA[s][t]}{\sigma}$ is not a \sigmaSE-extension of $\dfield{\AA[s]}{\sigma}$. Then there is a $\gamma\in\AA[s]$ with $\sigma(\gamma)-\gamma=\kappa_1\,\beta_1+\dots+\kappa_e\,\beta_e$. Let $j$ be maximal such that $\kappa_j$ is non-zero. Then we conclude that $\sigma(\gamma')-\gamma'=\beta_j$ with
$$\gamma':=\frac{1}{\kappa_j}(\gamma-\kappa_1\,t_1-\dots-\kappa_{j-1}\,t_{j-1})\in\AA[s][t_1]\dots[t_{j-1}]$$
which implies that $\dfield{\AA[s][t_1]\dots[t_j]}{\sigma}$ is not a \sigmaSE-extension of $\dfield{\AA[s][t_1]\dots[t_{j-1}]}{\sigma}$ by Theorem~\ref{Thm:SigmaExt}; a contradiction. Thus $\dfield{\AA[s][t]}{\sigma}$ is a \sigmaSE-extension of $\dfield{\AA[s]}{\sigma}$. Together with~\eqref{Equ:AgainSolInT} we can apply part~2 of Theorem~\ref{Thm:DRStatement}. This concludes the proof.~\qed
\end{proof}

\section{A refinement to the class of indefinite nested sums over mixed \hbox{($q$--)}hypergeometric products}\label{Sec:SequenceSetting}

In Theorems~\ref{Thm:DRStatement} and~\ref{Thm:DRStatementGeneral} we established criteria for the simplification of our double sum in the setting of difference rings. More precisely, we assumed that we are given a \sigmaSE-extension $\dfield{\AA[s]}{\sigma}$ of $\dfield{\AA}{\sigma}$ with $\sigma(s)=s+\sigma(x)$ for some fixed $x\in\AA$ and derived criteria when one can find a $g\in\AA[s]$ or in an appropriate \sigmaSE-extension such that $g$ solves the telescoping equation~\eqref{Equ:TeleDF} with $f=k\,x\,s$.
In the following we will transfer this result from the  difference ring $\dfield{\AA[s]}{\sigma}$ to the ring of sequences $\dfield{\seqK}{\Shift}$. 
To this end, we assume that we are given a ring embedding, i.e., an injective ring homomorphism $\tau$ from $\AA$ into $\seqK$ with the additional property that $\tau(\sigma(f))\equiv\Shift(\tau(f))$ holds for all $f\in\AA$, i.e., we require that the diagram
$$\xymatrix@C0.3cm@R=0.3cm{
\AA\ar[dd]^{\tau}\ar[rr]^{\sigma}&&\AA\ar[dd]^{\tau}\\
&&\\
\seqK\ar[rr]^{\Shift}&&\seqK
}$$
commutes. In addition, we assume naturally that $\tau(c)\equiv(c)_{n\geq0}$ holds for all $c\in\KK$. Such a map $\tau$ is also called a \emph{$\KK$-embedding} (it is called a \emph{$\KK$-homomorphism} if the injectivity of $\tau$ is dropped). Note that for such a $\KK$-embedding it follows that $\tau(\AA)$ is a subring of $\seqK$ and $\Shift$ restricted to $\tau(\AA)$ forms a ring automorphism. Note that $\dfield{\AA}{\sigma}$ and $\dfield{\tau(\AA)}{\Shift}$ are the same up to renaming of the elements by $\tau$.

\begin{psexample}\label{Exp:RatDR}
Consider the difference field $\dfield{\KK(k)}{\sigma}$ from Example~\ref{Exp:Q(k)}
with the evaluation function $\fct{\ev}{\KK(k)\times\NN}{\KK}$ as in~\eqref{Equ:EvalRat}. Then we can define the map $\fct{\tau}{\KK(k)}{\seqK}$ with $\tau(f)=(\ev(f,i))_{i\geq0}$
for $f\in\KK(k)$. One can easily see that $\tau$ is a ring homomorphism and with~\eqref{Equ:EvHomomShift} it follows that $\tau$ is a $\KK$-homomorphism. Finally, $\tau(f)\equiv0$ implies that $f=0$ since the numerator and denominator of $f$ can have only finitely many roots. Consequently, $\tau$ is a $\KK$-embedding. The subdifference ring $\dfield{\tau(\KK(k))}{\Shift}$ of $\dfield{\seqK}{\Shift}$ is also called the \emph{difference ring of rational sequences}. 
\end{psexample}

\begin{psexample}\label{Exp:EmbedHarmonicN}
Consider the \sigmaSE-extension $\dfield{\KK(k)[s]}{\sigma}$ of $\dfield{\KK(k)}{\sigma}$ from Ex.~\ref{Exp:K(k)[s]Sigma} (see also Ex.~\ref{Exp:Q(k)}) with the corresponding evaluation function $\fct{\ev}{\KK(k)[s]\times\NN}{\KK}$ that models the harmonic numbers $H_k$ with $s$. Then using similar arguments as in Example~\ref{Exp:RatDR} we conclude that $\fct{\tau}{\KK(k)[s]}{\seqK}$ defined by 
$\tau(f)=(\ev(f,i))_{i\geq0}$ for $f\in\KK(k)[s]$ is a $\KK$-homomorphism. By difference ring theory~\cite{DR3} it follows that $\tau$ is injective, and thus $\tau$ is a $\KK$-embedding.
\end{psexample}

More generally, we succeeded in such a construction in~\cite{DR3} not only for the harmonic numbers $H_k$ as elaborated in Example~\ref{Exp:EmbedHarmonicN} but for the general class of sequences that can be given in terms of nested sums over hypergeometric/$q$-hypergeometric/mixed-hypergeometric products. 

\begin{psdefinition}\label{Equ:DefNestedSums}
Let $\KK=\KK'(q_1,\dots,q_v)$ be a rational function field where $\KK'$ is a field of characteristic $0$. A product $\prod_{j=l}^kf(j,q_1^j,\dots,q_v^j)$, $l\in\NN$, is called \emph{mixed-multibasic hypergeometric}~\cite{Bauer:99} (in short \emph{mixed hypergeometric}) in $k$ over $\KK$ if $f(y,z_1,\dots,z_v)$ is an element from the rational function field $\KK(y,z_1,\dots,z_v)$ where the numerator and denominator of
$f(j,q_1^j,\dots,q_v^j)$ are nonzero for all $j\in\ZZ$ with $j\geq l$. Such a product is evaluated to a sequence following the rule
\begin{equation*}
\prod_{j=l}^{k} f(j,q_1^j,\dots,q_v^j): \ZZ\rightarrow \KK, m\mapsto
\begin{cases}
\prod_{j=l}^{m} f(j,q_1^j,\dots,q_v^j), &\mbox{if } l\leq m\\
1, & \mbox{otherwise}.
\end{cases}
\end{equation*}
Further, such a product is called \emph{$q$-hypergeometric} if $f$ is free of $y$, $v=1$ and $q_1=q$, i.e., $f\in\KK(z_1)$ with $\KK=\KK'(q)$. It is called \emph{hypergeometric} if $v=0$, i.e., $f\in\KK(y)$ with $\KK=\KK'$.\\
An \emph{expression in terms of nested sums over hypergeometric/$q$-hypergeometric/ mixed hypergeometric products in $k$ over $\KK$} is composed recursively by the three operations 
($+,-,\cdot$) with
\begin{itemize}
\item elements from the rational function field $\KK(k)$,
\item hypergeometric/$q$-hypergeometric/mixed hypergeometric products in $k$ over $\KK$,
\item and sums of the form $\sum_{j=l}^kf(j)$ with $l\in\NN$ where 
$f(j)$ is an expression in terms of nested sums over hypergeometric/$q$-hypergeometric/mixed hypergeometric products in $j$ over $\KK$; here it is assumed that the evaluation\footnote{Note that $\KK\subseteq\KK_X$ and thus the evaluation of a sum has been defined already in~\eqref{Equ:SumEv}.} of $f(j)|_{j\mapsto\lambda}$ for all $\lambda\in\ZZ$ with $\lambda\geq l$ does not introduce any poles.
\end{itemize}
\end{psdefinition}

Given such an expression $F(k)$ the evaluation $F(k)|_{k\mapsto\lambda}$ might be only defined for all $\lambda\geq l$ for some $l\in\NN$. In order to obtain an evaluation for all $\lambda\in\NN$, we set $F(k)|_{k\mapsto\lambda}=0$ for $\lambda=0,\dots,l-1$. 
Similarly to Definition~\ref{def:generic sums} we will give such products and sums defined over such products two different meanings. They form expressions that evaluate to sequences as introduced above, or they are just shorthand notations for the underlying sequences $(F(k)|_{k\mapsto\lambda})_{\lambda\geq0}$. The meaning (expression or sequence) of such a sums or products will be always clear from the context.
E.g., the harmonic numbers $H_n$ or the left- and right-hand sides of~\eqref{Equ:CalkinVar2a} 
and~\eqref{C2 Va} are either expressions in terms of indefinite nested sums over hypergeometric products in $a$ over $\KK=\QQ(n)$ or they are shorthand notations for sequences in $\KK$.

In general, as the sum $H_k\in\seqK$ can be rephrased in the difference ring $\dfield{\KK(k)[s]}{\sigma}$ given in Example~\ref{Exp:EmbedHarmonicN}, we can represent nested sums as defined in Definition~\ref{Equ:DefNestedSums} in a particular class of difference rings called 
\emph{simple \rpisiSE-rings}; for their definition we refer to the Appendix~\ref{Sec:Appendix}. At this point we want to emphasize
only the following crucial properties~\cite{DR3,DR4} of simple \rpisiSE-rings that enable one to treat the above class of nested sums in full generality.

\begin{pstheorem}\label{Thm:EmbedSigmaExt}
Let $\bar{X}_k(=\bar{X}(k))\in\seqK$ be a sequence given in terms of nested sums over hypergeometric (resp.\ $q$-hyper\-geometric or mixed hypergeometric) products where $\KK$ is algebraically closed\footnote{Algorithmically, one starts with a base field $K$ (like $\QQ$ or $\QQ(n)$) and constructs ---if necessary--- a finite algebraic extension of it such that statement (1) is true.}. Then the following holds.

\vspace*{-0.2cm}

\begin{enumerate}
\item[(1)] There is a simple \rpisiSE-ring $\dfield{\AA}{\sigma}$ with constant field $\KK$ equipped with a $\KK$-embedding $\fct{\tau}{\AA}{\seqK}$ and with $x\in\AA$ such that $\tau(x)\equiv \bar{X}_k$ holds.
\end{enumerate}

\vspace*{-0.2cm}

Moreover, for this $\tau$ one has:

\vspace*{-0.2cm}

\begin{enumerate}
\item[(2a)] For any $h\in\AA$ there is a sequence $H(k)$ expressible in terms of nested sums over hypergeometric (resp.\ $q$-hyper\-geometric or mixed hypergeometric) products with $\tau(h)\equiv H(k)$.
\item[(2b)] If the difference ring extension $\dfield{\AA[s]}{\sigma}$ of $\dfield{\AA}{\sigma}$ with $s$ being transcendental over $\AA$ and  $\sigma(s)=s+\sigma(x)$, $x$ as in part (1), forms a \sigmaSE-extension, then the difference ring homomorphism $\fct{\tau'}{\AA[s]}{\seqK}$ defined by $\tau'|_{\AA}=\tau$ and $\tau'(s)\equiv\sum_{k=0}^n \bar{X}_k$ forms a $\KK$-embedding\footnote{This means that $\tau(\sum_{i=0}^r f_i\,s^i)\equiv\sum_{i=0}^r\tau(f_i)\big(\big(\sum_{k=0}^n \bar{X}_k\big)^i\big)_{n\geq0}$ for $f_0,\dots,f_r\in\AA$.}. 
\end{enumerate}
In particular, the simple \rpisiSE-ring $\dfield{\AA}{\sigma}$ with $f$ and the embedding $\tau$ can be computed explicitly; for further details see Appendix~\ref{Sec:Appendix}. 
\end{pstheorem}

\noindent 
Note that part~(1) implies that a finite number of nested sums over hypergeometric, $q$-hyper\-geometric or mixed hypergeometric can be always formalized in a simple \rpisiSE-ring, and part~(2a) states that any element in such a ring can be reinterpreted as such a sum or product.
This representation justifies the following definition.

\begin{psdefinition}
A sub-difference ring $\dfield{\SubS}{\Shift}$ of $\dfield{\seqK}{\Shift}$ is called a \emph{product-sum sequence ring}, if there is a simple \rpisiSE-ring $\dfield{\AA}{\sigma}$ with constant field $\KK$ together with a $\KK$-embedding $\fct{\tau}{\AA}{\seqK}$ with $\tau(\AA)=\SubS$.
\end{psdefinition}

Now let us reconsider our difference ring calculations of Subsection~\ref{Sec:DRAlg} within such
a product-sum sequence ring $\dfield{\SubS}{\Shift}$ where $\bar{X}_k$ stands for a sequence that is given in terms of nested sums over products. According to Theorem~\ref{Thm:EmbedSigmaExt}, this means that  
there is a simple \rpisiSE-ring $\dfield{\AA}{\sigma}$ with constant field $\KK$ equipped with a $\KK$-embedding $\fct{\tau}{\AA}{\seqK}$ and with an $x\in\AA$ such that $\tau(x)\equiv \bar{X}_k$ holds. Suppose the decision procedure implemented in \texttt{Sigma} tells us (as above in Example~\ref{Exp:K(k)[s]Sigma}) that there is no $g\in\AA$ such that $\sigma(g)=g+\sigma(x)$ holds. Note that this implies that there is no sequence $G(k)\in\tau(\AA)$ expressible in terms of nested sums with $G(k+1)-G(k)\equiv\bar{X}_{k+1}$ or equivalently it follows that 

\vspace*{-0.3cm}

$$\sum_{i=0}^k\bar{X}_i\notin\tau(\AA).$$

\vspace*{-0.1cm}

\noindent Furthermore, we conclude by part~(2b) of Theorem~\ref{Thm:EmbedSigmaExt} that we can extend the $\KK$-embedding $\tau$ from $\AA$ to $\AA[s]$ with 
$\tau(s)\equiv\sum_{i=0}^k\bar{X}_k.$
From this it can be derived that $\dfield{\AA[s]}{\sigma}$ and $\dfield{\tau(\AA[s])}{\Shift}$ are isomorphic, i.e., the difference rings are the same up to renaming of the objects using $\tau$.

\noindent With this background we restart our calculations to obtain a solution $g$ of the telescoping equation

\vspace*{-0.3cm}

\begin{equation}\label{Equ:TeleCalk2SimpleDR}
\sigma(g)-g=(k+1)\,\sigma(x\,s)=(k+1)\sigma(x)(s+\sigma(x)).
\end{equation}

\vspace*{-0.1cm}

\noindent In the first major step we assumed that we can find a $c\in\KK$ and a $y\in\AA$ such that~\eqref{Equ:ConstraintFory} holds. Now let $\bar{Y}_k$ be the sequence in terms of nested sums with $\tau(y)\equiv \bar{Y}_k\in\tau(\AA)$. Then by construction it follows that~\eqref{Equ:PTele} holds for $\bar{Y}_k$ and $c$. 

We proceed with our calculations by entering in the already worked out case distinction.\\
\textit{Case 1.} We can compute a $d\in\KK$ and $g_0\in\AA$ with~\eqref{Deg0DR}. 
Then for the sequence $G_0(k)$ with $\tau(g_0)=G_0(k)$ in terms of nested sums we obtain
\begin{equation}\label{Equ:G0SeqConstr}
G_0(k+1)-G_0(k)\equiv\bar{X}_{k+1}^2-c \bar{X}_{k+1}^2+k \bar{X}_{k+1}^2-\bar{X}_{k+1} \bar{Y}_{k+1}-d\,\bar{X}_{k+1}.
\end{equation}
Further, the $g\in\AA[s]$ with~\eqref{Equ:gCase1} is a solution of~\eqref{Equ:TeleCalk2SimpleDR} 
under the assumption that $c\in\KK$ and $y$ are a solution of~\eqref{Equ:ConstraintFory}. This implies that

\vspace*{-0.4cm}

$$\Shift(\tau(g))-\tau(g)\equiv\tau((k+1)\,\sigma(x)\,(s+\sigma(x))\equiv((k+1)\,\bar{X}_{k+1}\,(\sum_{i=0}^k\bar{X}_i+\bar{X}_{k+1}))_{k\geq0}.$$

\vspace*{-0.3cm}

\noindent By construction, we obtain
$\tau(g)\equiv G(k)\in\tau(\AA[s])$ with 
$G(k)=c\,\Big(\sum_{i=0}^k\,\bar{X}_i\Big)^2+(\bar{Y}_k+d)\,\sum_{i=0}^k\,\bar{X}_i+G_0(k)$,
and thus $G(k)$ is a solution of
\begin{equation}\label{Equ:TeleSeqCalkinVar1}
G(k+1)-G(k)\equiv(k+1)\,\bar{X}_{k+1}\Big(\sum_{j=0}^k\bar{X}_j+\bar{X}_{k+1}\Big)
\end{equation}
under the constraint that~\eqref{Equ:PTele} holds for $\bar{Y}_k$ and $c\in\KK$.
Passing from the generic sequence variable
$k$ to concrete integers $k\in \ZZ$, using 
\eqref{Equ:TeleSeqCalkinVar1} we can check that telescoping yields
\begin{equation}\label{Equ:Calkin2VarG0Summable}
\sum_{k=0}^ak\,\bar{X}_k\sum_{j=0}^k\bar{X}_j=G(a)-G(-1)=
c\,\Big(\sum_{i=0}^a\,\bar{X}_i\Big)^2+(\bar{Y}_a+d)\,\sum_{i=0}^a\,\bar{X}_i+G_0(a)-G_0(-1).
\end{equation}
\textit{Case~2.} There does not exist a $d\in\KK$ and $g_0\in\AA$ with~\eqref{Deg0DR}. By Theorem~\ref{Thm:EmbedSigmaExt} we can extend the $\KK$-embedding from $\AA[s]$ to $\AA[s][t]$ with $\tau(t)\equiv G_0(k)$ where 

\vspace*{-0.3cm}

\begin{equation}\label{Equ:G0Sum}
G_0(k)=\sum_{i=0}^k(-c \bar{X}_{i}^2+i \bar{X}_{i}^2-\bar{X}_{i} \bar{Y}_{i}).
\end{equation} 

\vspace*{-0.1cm}

\noindent In particular, we conclude that $G_0(k)\notin\tau(\AA)$. Moreover, the solution~\eqref{Equ:gCase2} of~\eqref{Equ:TeleCalk2SimpleDR} yields the solution~\eqref{Equ:Calk2VarTeleSol} of~\eqref{Equ:TeleSeqCalkinVar1} 
under the constraint that~\eqref{Equ:PTele} holds for $\bar{Y}_k$ and $c\in\KK$. Finally, we arrive at our simplification given in~\eqref{Equ:CalkinVarOneSum2Ev}.
\medskip

In Theorem~\ref{Thm:DRStatement} of Section~\ref{Sec:DRAlg} we summarized the considerations leading to cases (1) and (2).
Before we can reformulate these cases in the context of sequences, we collect some key properties indicated already above.

\begin{pslemma}\label{Lemma:RephrasetoSequ}
Let $\dfield{\AA}{\sigma}$ be a simple \rpisiSE-ring (see Definition~\ref{Def:SimpleRing}) with constant field $\KK$, and let
$\fct{\tau}{\AA}{\seqK}$ be a $\KK$-embedding. Set $\SubS=\tau(\AA)$ and let $f\in\AA$ with $\tau(f)\equiv F= (F(k))_{k\geq0}\in\SubS$ and define $\bar{S}:=(\sum_{j=0}^kF(j))_{k\geq0}\in\seqK$. Then the following statements are equivalent.
\begin{myEnumerate}
\item There is a \sigmaSE-extension $\dfield{\AA[s]}{\sigma}$ of $\dfield{\AA}{\sigma}$ with $\sigma(s)=s+\sigma(f)$.
\item There is no $G\in\SubS$ with $\Shift(G)-G\equiv \Shift(F)$.
\item $\SubS[\bar{S}]$ forms a polynomial ring.
\item $\bar{S}\notin\SubS$.
\end{myEnumerate}
\end{pslemma}
\begin{proof}
$(1)\Leftrightarrow(2)$: 
There is a \sigmaSE-extension $\dfield{\AA[s]}{\sigma}$ of $\dfield{\AA}{\sigma}$ iff there is no $g\in\AA$ with $\sigma(g)=g+\sigma(f)$ by Theorem~\ref{Thm:SigmaExt}. Since $\tau$ is a $\KK$-embedding, the latter condition is equivalent to saying that there is no $G\in\tau(\AA)$ with $\Shift(G)-G\equiv\tau(\sigma(f))\equiv\Shift(\tau(f))\equiv\Shift(F)$.\\
$(1)\Rightarrow(3)$: By part~(2b) of Theorem~\ref{Thm:EmbedSigmaExt} one can extend $\tau$ from $\AA$ to $\AA[s]$ by $\tau(s)\equiv S$. Since $\AA[s]$ is a polynomial ring, $\SubS[\bar{S}]$ forms a polynomial ring.\\ 
$(3)\Rightarrow(4)$ holds trivially.\\
$(4)\Rightarrow(2)$: Suppose that there is a $G\in\SubS$ with $\Shift(G)-G\equiv\tau(\sigma(f))$. Since $\Shift(\bar{S})\equiv\bar{S}+(F(k+1))_{k\geq0}\equiv\bar{S}+(F(k))_{k\geq1}\equiv\bar{S}+\Shift(\tau(f))\equiv\bar{S}+\tau(\sigma(f))$, we conclude that $\Shift(\bar{S}-G)\equiv\bar{S}-G$ and thus $\bar{S}\equiv G+(c,c,c,\dots)$ for some $c\in\KK$. Hence $\bar{S}\in\SubS$.~\qed
\end{proof}

\noindent With Lemma~\ref{Lemma:RephrasetoSequ} and the above considerations the statements of part 1 of Theorem~\ref{Thm:DRStatement} and Theorem~\ref{Thm:DRStatementGeneral} (which is a slightly more general version of part~2 of Theorem~\ref{Thm:DRStatement}) translate directly to the corresponding statements of the following Theorem~\ref{Thm:SeqStatement}. 

\begin{pstheorem}\label{Thm:SeqStatement}
Let $\dfield{\SubS}{\Shift}$ be a product-sum sequence ring containing the sequence $k$ with $\Shift(k)=k+1$.
Let $\bar{X}_k\in\SubS$ and suppose that $\sum_{i=0}^k \bar{X}_i\notin\SubS$. Then within the polynomial ring $\SubS':=\SubS[\sum_{i=0}^k \bar{X}_i]$ the following two statements hold:
\begin{myEnumerate}
\item $\sum_{k=0}^ak\,\bar{X}_k\sum_{i=0}^k\bar{X}_i\in\SubS'$ iff
\begin{enumerate}
 \item[(a)] there is a $\bar{Y}_k\in\SubS$ and $c\in\KK$ with~\eqref{Equ:PTele},
 \item[(b)] and there is a $G_0(k)\in\SubS$ and $d\in\KK$ with~\eqref{Equ:G0SeqConstr} 
 (where $c$ is the one from part (a)).
\end{enumerate}
If (a) and (b) hold, we get the simplification given in~\eqref{Equ:Calkin2VarG0Summable}. 

\vspace*{0.2cm}

\item Suppose that $Z_a:=\sum_{k=0}^ak\,\bar{X}_k\sum_{i=0}^k\bar{X}_i\notin\SubS'$. Then the sequence $Z_a$ can be given in terms of single nested sums whose summands are from $\SubS$ iff the following two statements hold:

\vspace*{-0.2cm}

\begin{enumerate}
\item[(a)] there is a $\bar{Y}_k\in\SubS$ and $c\in\KK$ with~\eqref{Equ:PTele},
\item[(b)] there is no $G_0(k)\in\SubS$ and $d\in\KK$ with~\eqref{Equ:G0SeqConstr} (where $c$ is the one from part (a)).
\end{enumerate}
If (a) and (b) hold, we obtain the simplification~\eqref{Equ:CalkinVarOneSum2}. 
\end{myEnumerate}
\end{pstheorem}


\section{Using the \texttt{Sigma} package}\label{Sec:Sigma}

\subsection{The symbolic approach with \texttt{Sigma}}\label{Sec:SymbolicApproach}

As already demonstrated in~\myIn{\ref{MMA:SigmaReduceXList}} the difference ring machinery is activated in \texttt{Sigma} by executing the function call \texttt{SigmaReduce} to the given summation problem. If a generic sequence $X_k$ arises within the summation problem, this information has to be passed to \texttt{SigmaReduce} with the option \texttt{XList}$\to\{X\}$. Then the generic sequence $X_k$ and its shifted versions $\dots,X_{k-2},X_{k-1},X_k,X_{k+1},X_{k+2},\dots$ are represented by the variables $\dots,x_{-2},x_{-1},x_0,x_1,x_2,\dots$, respectively. Namely, as worked out in~\cite{KS:06a,KS:06b} \texttt{Sigma} takes the field $\GG=\KK(\dots,x_{-2},x_{-1},x_0,x_1,x_2,\dots)$ with infinitely many variables and uses the field automorphism $\fct{\sigma}{\GG}{\GG}$ with $\sigma(x_i)=x_{i+1}$ for all $i\in\ZZ$ and $\sigma(c)=c$ for all $c\in\KK$. 
The obtained difference field $\dfield{\GG}{\sigma}$ with $\const{\GG}{\sigma}=\KK$ is also called the \emph{difference field of free sequences}. In order to define the underlying evaluation function for $\GG$, the constant field $\KK$ has to be constructed accordingly. Here one takes the rational function
field $\KK=\KK'(\dots,X_{-2},X_{-1},X_0,X_1,X_2,\dots)$ again with infinitely many variables where $\KK'$ is a field of characteristic $0$; note that $\KK'_X$ (see our earlier Definition~\ref{Equ:KXDef}) and $\KK$ are closely related: $\KK'_X$ is the polynomial ring in the variables $X_i$ with $i\in\ZZ$ and $\KK$ is simply its quotient field. The evaluation function $\ev$ for $\GG$ is provided with $\ev(x_i,j)=X_{i+j}$ for $i,j\in\ZZ$.\\
Usually, in generic summation problems as considered in this article, the summation input of \texttt{SigmaReduce} depends not only on generic sequences, but on generic sums (see Definition~\ref{def:generic sums}) and more generally, on nested sums and products defined over generic sequences. In this case, the input expression is represented accordingly with a tower of \rpisiSE-extensions over $\dfield{\GG}{\sigma}$, see the Appendix~\ref{Sec:Appendix}, which leads to a difference ring $\dfield{\AA}{\sigma}$. This construction can be carried out automatically by the tools given in~\cite{DR1,DR2,DR3} in combination with the machinery described in~\cite{KS:06a,KS:06b}. Finally, \texttt{Sigma} tries to simplify the given summation problem using the different telescoping algorithms from~\cite{FastAlgorithm1,FastAlgorithm2,FastAlgorithm3}.

\medskip

\noindent\textit{Calculation steps for Subsection~\ref{Sec:Calkin1}:} In order to tackle the sum on the left-hand side of~\eqref{Equ:Calkin1Simple} \texttt{Sigma} represents $X_j$ by $x_0\in\GG$. By default the difference field extension $\dfield{\GG(k)}{\sigma}$ of $\dfield{\GG}{\sigma}$ with $\sigma(k)=k+1$ and $\const{\GG(k)}{\sigma}=\KK$ is adjoined automatically. Furthermore, the \sigmaSE-extension $\dfield{\GG(k)[s]}{\sigma}$ of $\dfield{\GG(k)}{\sigma}$ with $\sigma(s)=s+x_1$ is constructed to model the generic sum $\sum_{j=0}^kX_j$ with $\sum_{j=0}^{k+1}X_j=\sum_{j=0}^kX_j+X_{k+1}$; internally Theorem~\ref{Thm:SigmaExt} is applied to check that this is indeed a \sigmaSE-extension. As a consequence, we have that $\const{\GG(k)[s]}{\sigma}=\KK$. Now exactly the steps from Subsection~\ref{Sec:Calkin1} with $f=\sigma(s)=s+x_1$ are carried out in this difference ring, and the expression~\eqref{Equ:Calkin1Simple} (with the options \texttt{SimpleSumRepresentation}$\to$\texttt{True} and \texttt{SimplifyByExt$\to$MinDepth} activated; see Remark~\ref{Remark:XYCalkin1} for further explanations) is returned.

\medskip

\noindent\textit{Calculation steps for Subsection~\ref{Sec:Calkin2Simple}:}
The tactic of Subsection~\ref{Sec:Calkin1} fails for the double sum on the left-hand side of~\eqref{Equ:CalkinVarOneSum2}. But, using in addition the \texttt{Sigma}-option
\texttt{ExtractConstraints}$\to\{Y\}$, as demonstrated in~\myIn{\ref{MMA:SigmaReduceNotSimpleWithConstraints}}, the new machinery introduced in Section~\ref{Sec:DRAlg} is activated. Internally, again the difference ring $\dfield{\GG(k)[s]}{\sigma}$ with constant field $\KK$ is constructed, and the computation steps are carried out with $\sigma(f)=(k+1)x_1(s+x_1)$ (instead of $\sigma(f)=(k+1)\sigma(x)(s+\sigma(x))$. They are precisely the same as in Section~\ref{Sec:DRAlg}. In this process we produce the constraint
$$\sigma(g_1)-g_1=(1+k)x_1-2\,c\,x_1;$$
compare with~\eqref{Equ:ConstraintDeg1DR}.
Since \texttt{Sigma} does not find a solution $g_1\in\GG(k)[s]$, it extends the underlying difference field $\GG$ by the new variables $\dots,y_{-2},y_{-1},y_0,y_1,y_2,\dots$ and extends the automorphism $\sigma$ with $\sigma(y_i)=y_{i+1}$ for all $i\in\ZZ$. Now we continue our calculation with $g_1=y_i+d$ and a new variable $c$ (i.e., we extend the constant field $\KK$ by $c$) and obtain the constraint
$$\sigma(g_0)-g_0=x_1^2-c x_1^2+k x_1^2-x_1 y_1-d\,x_1$$
of $g_0$; compare with~\eqref{Deg0DR}. Since we do not find a $g_0\in\GG(k)(s)$ (with the updated $\GG$ containing now also the variables $y_i$ with $i\in\ZZ$ and the new constant $c$) and $d\in\KK(c)$, we construct the \sigmaSE-extension $\dfield{\GG(k)[s][t]}{\sigma}$ of $\dfield{\GG(k)[s]}{\sigma}$ with 
$$\sigma(t)=t+(x_1^2-c x_1^2+k x_1^2-x_1 y_1).$$
This finally produces the solution $g=c\,s^2+y_0\,s+t$. Reinterpreting this result in terms of the generic sequences $X_k$ and $Y_k$ produces the output~\myOut{\ref{MMA:SigmaReduceNotSimpleWithConstraints}}.

\medskip

Concerning this concrete summation problem the following remarks are relevant.
\begin{enumerate}
 \item The output~\myOut{\ref{MMA:SigmaReduceNotSimpleWithConstraints}} provides the full information that is needed to apply Theorem~\ref{Thm:SeqStatement} taking care of the two possible scenarios. Specializing $X_k$ and $Y_k$ (where $Y_k$ and $c$ are solutions of the constraint~\eqref{Equ:PTele}) to concrete sequences in $\dfield{\SubS}{\Shift}$, it might happen that the found sum extension simplifies further in the given ring $\SubS$. This situation is covered by part~(1) of Theorem~\ref{Thm:SeqStatement}. Otherwise, if the sum cannot be simplified in~$\SubS$, part~(2) of the Theorem~\ref{Thm:SeqStatement} can be applied.     
\item Fix a product-sum sequence ring $\dfield{\SubS}{\Shift}$. If $\sum_{j=0}^k\bar{X}_j\notin\SubS$, the output gives a full characterization when the sum $\sum_{k=0}^a\bar{X}_k\sum_{j=0}^k\bar{X}_j$ can be written as an expression in terms of single nested sums; see Theorem~\ref{Thm:SeqStatement} for further details. However, if we enter the special case $\sum_{j=0}^k\bar{X}_j\in\SubS$, then the result provides only a sufficient criterion to get such a simplification. Still the toolbox can be applied also in such a case as worked out in Example~\ref{Exp:Calkin S1}; there we chose $X_j=H_j$ for which the simplification $\sum_{j=0}^k\bar{X}_j=-n + (1 + n) H_n$ is possible.

\item Specializing the identities in~\eqref{Equ:Calkin1Simple} to concrete sequences $\bar{X}_k$ often leads to further simplifications.
\end{enumerate}

We considered the very special case of the input expression $\sum_{k=0}^ak\,X_k\sum_{i=0}^kX_i$. However, the proposed method works for any input sum $\sum_{k=0}^af(k)$  where the summand $f(k)$ is built by a finite number of generic sequences, say  $X,Y,\dots,Z$, and over nested sums over hypergeometric/$q$-hypergeometric/mixed hypergeometric products. A typical function call, for instance, is~\myIn{\ref{MMA:XYExp1}}. Here the same ideas are applied as in Section~\ref{Sec:BasicTactics} where instead of $\sum_{i=0}^kX_i$ the most nested sum (and among the most nested sums the one with highest degree) of the summand $f(k)$ is chosen. In particular, the following refinements can be activated.

\begin{enumerate}
\item In Subsection~\ref{Sec:Calkin2Simple} we combined 
the telescoping algorithm from~\cite{FastAlgorithm2} with our new idea to extract constraints in form of parameterized telescoping equations and to encode these constraints in the output expression by using new generic sequences. Within \texttt{Sigma} also other enhanced telescoping strategies for simplification ~\cite{Schneider:07d,FastAlgorithm2,FastAlgorithm3} can be combined with this new feature. For further details on the possible options we refer also to Remarks~\ref{Remark:XYCalkin1} and~\ref{Remark:XYCalkin2}.

\item In Subsection~\ref{Sec:Calkin2Simple} the most complicated sum occurs only linearly. As a consequence we run into three constraints given by step-wise coefficient comparison. Namely, for our ansatz~\eqref{Equ:AnsatzG3} we get the constraint~\eqref{Equ:Degree2}, which can always be treated, the constraint~\eqref{Equ:Tele Y1} where we introduced a generic sequence $Y_k$ subject to the parameterized telescoping relation~\eqref{Equ:Tele Y1 sequence version}, and the constraint~\eqref{Equ:Tele5} which we could handle by the sum extension~\eqref{Equ:G0Sum}. More generally, if the most complicated sum occurs with degree $d>1$, one ends up with $d+2$ constraints. Some of them can be solved directly by \texttt{Sigma} within the given difference ring, but in general there will remain constraints which can only be treated by introducing a new generic sequence that must satisfy a certain parameterized telescoping equation. Activating the option \texttt{ExtractConstraints}$\to\{Y^{(1)},\dots,Y^{(l)}\}$, \texttt{SigmaReduce} is allowed to provide (if necessary) up to $l$ constraints in form of parameterized telescoping equations, each one with a different generic sequence from $Y^{(1)},\dots,Y^{(l)}$. If not successful, i.e., if more than $l$ generic sequences are needed, \texttt{Sigma} gives up and returns the input expression.
\end{enumerate}

\subsection{Discovery of identities}\label{Sec:Discovery}

We illustrate how the presented techniques can support the (re)discovery of numerous identities. We start with the generic sum 

\begin{mma}
 \In mySum=\sum_{k=0}^a\Big(\sum_{j=0}^kX[j]\Big)^2;\\
\end{mma}

\noindent and obtain the following general simplification formula

 \begin{mma}
 \In \{closedForm,constraint\}=SigmaReduce[mySum,XList\to\{X\},ExtractConstraints\to\{Y\},\newline
 \hspace*{3cm}SimpleSumRepresentation\to False,RefinedForwardShift\to False]\\
 \Out \{(a
+c
) \big(
        \sum_{i=0}^a X[i]\big)^2
+
\sum_{i=0}^a \big(
        X[i]^2
        -c X[i]^2
        -i X[i]^2
        -X[i] Y[i]
\big)
+
Y[a]\sum_{i=0}^a X[i],\newline
\{Y[{a+1}]-Y[{a}]==-2 a X[a+1] -2\,c X[{a+1}]\}\}\\
\end{mma}

\noindent The result can be simplified further to the form

\begin{mma}
\In SigmaReduce[closedForm, a, XList\to\{X,Y\}, SimpleSumRepresentation\to True]\\
\Out (a
+c
) \big(
        \sum_{i=0}^a X[i]\big)^2
-c 
\sum_{i=0}^a X[i]^2
-
\sum_{i=0}^a X[i] Y[i]
+Y[a]\sum_{i=0}^a X[i]
+\sum_{i=0}^a X[i]^2-
\sum_{i=0}^a i X[i]^2\\
\end{mma}

\medskip

\noindent This means that the identity  
\small
\begin{equation}\label{Equ:GenericIdCalkin2}
\sum_{k=0}^a\Big(\sum_{j=0}^k\bar{X}_j\Big)^2=(a
+c
) \big(
        \sum_{k=0}^a \bar{X}_k\big)^2
-c 
\sum_{k=0}^a \bar{X}_k^2
-
\sum_{k=0}^a \bar{X}_k \bar{Y}_k
+\bar{Y}_a\sum_{k=0}^a \bar{X}_k
+\sum_{k=0}^a \bar{X}_k^2-
\sum_{k=0}^a k \bar{X}_k^2
\end{equation}
\normalsize
holds for any sequences $(\bar{X}_k)_{k\geq0}$, $(\bar{Y}_k)_{k\geq0}$ with $\bar{X}_k,\bar{Y}_k\in\KK$ and $c\in\KK$ if $c$ and $\bar{Y}_k$ are a solution of the parameterized telescoping equation
\begin{equation}\label{Equ:Calkin2Constr}
\bar{Y}_{k+1}-\bar{Y}_{k}=-2 k \bar{X}_{k+1} -2\,c\, \bar{X}_{k+1}.
\end{equation}
Even more holds by a straightforward variant of Theorem~\ref{Thm:SeqStatement}: if one takes a product-sum sequence ring $\dfield{\SubS}{\Shift}$ and takes a sequence $\bar{X}_k$ which is in $\SubS$ but where the sequence of $\sum_{j=0}^k\bar{X}_j$ is not in $\SubS$, then the double sum on the left-hand side of~\eqref{Equ:GenericIdCalkin2} can be simplified to single nested sums defined over $\SubS$ if and only if there is a solution $c\in\KK$ and $\bar{Y}_k$ in $\SubS$ of~\eqref{Equ:Calkin2Constr}. In this case the right-hand side of~\eqref{Equ:Calkin2Constr} with the explicitly given $c$ and $\bar{Y}_k$ produces such a simplification. 

\begin{psexample}
$\bar{X}_k=\binom{n}{k}$: Plugging the solution
$c=\frac{2-n}{2}$ and $\bar{Y}_k=\binom{n}{k}(-k+n)$ of~\eqref{Equ:Calkin2Constr} into~\eqref{Equ:GenericIdCalkin2} yields
\begin{align*}
\sum_{k=0}^a\Big(\sum_{j=0}^k&\binom{n}{j}\Big)^2=(-a
+n
)\binom{n}{a}  
\sum_{k=0}^a \binom{n}{k}
+\big(
        a
        +\frac{2-n}{2}
\big)
\big(
        \sum_{k=0}^a \binom{n}{k}\big)^2\\
&+
\sum_{k=0}^a \binom{n}{k}^2
-\frac{2-n}{2} 
\sum_{k=0}^a \binom{n}{k}^2
-
\sum_{k=0}^a k \binom{n}{k}^2
-
\sum_{k=0}^a \binom{n}{k}^2 (-k
+n
)\\
\stackrel{\texttt{Sigma}}{=}&\binom{n}{a} (-a
+n
) 
\sum_{k=0}^a \binom{n}{k}
+\frac{1}{2} (2
+2 a
-n
) \big(
        \sum_{k=0}^a \binom{n}{k}\big)^2
-\frac{1}{2} n 
\sum_{k=0}^a \binom{n}{k}^2
\end{align*}
which is valid for all $a,n\in\NN$.
Following the same tactic, we ``discover'' the identities
\begin{align*}
\sum_{k=0}^a \Big(
        \sum_{j=0}^k x^j \binom{n}{j}\Big)^2=&-\frac{nx}{x+1}\sum_{k=0}^a x^{2 k} \binom{n}{k}^2
+\tfrac{(1
+a
+x
+a x
-n x
)}{x+1}\Big(
        \sum_{k=0}^a x^k \binom{n}{k}\Big)^2\\
&+\frac{x-1}{x+1}\sum_{k=0}^a k x^{2 k} \binom{n}{k}^2
-\frac{2 x^{a+1}(a
-n
)}{x+1}\binom{n}{a}
\sum_{k=0}^a x^k \binom{n}{k},\\
\sum_{k=0}^a \Big(
        \sum_{j=0}^k (-1)^j \binom{n}{j}\Big)^2=&\frac{n}{2 (2 n-1)}\sum_{k=0}^a \binom{n}{k}^2
-\frac{(2 a-3 n+2)(a-n)^2}{2 n^2 (2 n-1)}\binom{n}{a}^2;
\end{align*}
the first identity holds for $x\in\KK\setminus\{-1\}$ and $a,n\in\NN$ and the second holds for $a,n\in\NN$ with $n\neq0$. Furthermore we obtain
\begin{align*}
\sum_{k=0}^a \Big(
        \sum_{j=0}^k \frac{x^j}{\binom{n}{j}}\Big)^2=&\frac{1
+n
+x}{x+1}\sum_{k=0}^a \frac{x^{2 k}}{\binom{n}{k}^2}
+\frac{x-1}{x+1}\sum_{k=0}^a \frac{k x^{2 k}}{\binom{n}{k}^2}\\
&+\frac{a
-n
+2 x
+a x
}{x+1}\Big(
        \sum_{k=0}^a \frac{x^k}{\binom{n}{k}}\Big)^2
-\frac{2 (a+1) x^{a+1}}{(x+1) \binom{n}{a}}\sum_{k=0}^a \frac{x^k}{\binom{n}{k}},\\
\sum_{k=0}^a \Big(
        \sum_{j=0}^k \frac{(-1)^j}{\binom{n}{j}}\Big)^2=&
\tfrac{(n+1)^2 (4 a n^2+22 a
   n+30 a+3 n^2+23 n+38)
}{2 (n+2)^2 (n+3) (2 n+5)}        
+\tfrac{2 (-1)^a (a+1) (a+2) (n+1)}{(n+2)^2 (n+3)}\frac1{\binom{n}{a}}\\
&+\frac{(a+1)^2 (6
+2 a
+n
)}{2 (n+2)^2 (2 n+5)}\frac1{\binom{n}{a}^2}
+\frac{n+2}{2 (2 n+5)}\sum_{k=0}^a \frac{1}{\binom{n}{k}^2}
\end{align*}
for all $x\in\KK\setminus\{-1\}$ and $a,n\in\NN$ with $a\leq n$.
\end{psexample}

\medskip

Similarly, for the generic double sum

\begin{mma}
\In mySum=
\sum_{k=0}^a (-1)^k \Big(
        \sum_{j=0}^k X[j]\Big)^2;\\
\end{mma}

\noindent \texttt{Sigma} finds the general simplification

\begin{mma}
 \In \{closedForm,constraint\}=SigmaReduce[mySum,XList\to\{X\},ExtractConstraints\to\{Y\},\newline
 \hspace*{3cm}SimpleSumRepresentation\to False,RefinedForwardShift\to False]\\
 \Out \{-\frac{1}{2} c \big(
        \sum_{i=0}^a X[i]\big)^2
+\frac{1}{2} (-1)^a \big(
        \sum_{i=0}^a X[i]\big)^2
+\frac{1}{2} 
\sum_{i=0}^a \big(
        (-1)^i X[i]
        +c X[i]
        +Y[i]
\big) X[i]
-\frac{1}{2} 
Y[a]\sum_{i=0}^a X[i]
,
\newline
\{Y[a+1]-Y[a]==2 (-1)^a X[a+1]-2\,c X[a+1]\}\}.\\
\end{mma}

\noindent where the result can be simplified further to

\begin{mma}
\In SigmaReduce[closedForm, a, XList\to\{X,Y\}, SimpleSumRepresentation\to True]\\
\Out\big(
        -\frac{c}{2}
        +\frac{1}{2} (-1)^a
\big)
\big(
        \sum_{i=0}^a X[i]\big)^2
+\frac{1}{2} c 
\sum_{i=0}^a X[i]^2
+\frac{1}{2} 
\sum_{i=0}^a (-1)^i X[i]^2
+\frac{1}{2} 
\sum_{i=0}^a X[i] Y[i]
-\frac{1}{2}Y[a] 
\sum_{i=0}^a X[i]
 \\
\end{mma}

\medskip

\noindent This means that for any sequences $\bar{X}_k\in\KK$, $\bar{Y}_k\in\KK$ and $c\in\KK$ with \begin{equation}\label{Equ:ConstrainAltCalk2}
\bar{Y}_{k+1}-\bar{Y}_k=2 (-1)^k \bar{X}_{k+1}-2\,c\,\bar{X}_{k+1},
\end{equation}
we obtain the simplification
\begin{multline}\label{Equ:AltCalkin}
\sum_{k=0}^a (-1)^k \Big(
        \sum_{j=0}^k \bar{X}_j\Big)^2=
        \big(
        -\frac{c}{2}
        +\frac{1}{2} (-1)^a
\big)
\big(
        \sum_{k=0}^a \bar{X}_k\big)^2\\
+\frac{1}{2} c 
\sum_{k=0}^a \bar{X}_k^2
+\frac{1}{2} 
\sum_{k=0}^a (-1)^k \bar{X}_k^2
+\frac{1}{2} 
\sum_{k=0}^a \bar{X}_k \bar{Y}_k
-\frac{1}{2}\bar{Y}_a 
\sum_{k=0}^a \bar{X}_k.
\end{multline}
In addition, by a slight modification of Theorem~\ref{Thm:SeqStatement} we obtain the following stronger statement for any product-sum sequence ring $\dfield{\SubS}{\Shift}$ under the assumption that $\bar{X}_k$ is in $\SubS$, but $\sum_{j=0}^k\bar{X}_j$ is not in $\SubS$:
the double sum can be simplified to single nested sums defined over $\SubS$ 
if and only if~\eqref{Equ:AltCalkin} holds and there are $\bar{Y}_k\in\SubS$ and $c\in\KK$ with~\eqref{Equ:ConstrainAltCalk2}.

\noindent Again proceeding as above one can find, for instance, the following identities:
\begin{align*}
\sum_{k=0}^a (-1)^k \Big(
        \sum_{j=0}^k \binom{n}{j}\Big)^2=&
\frac{(-a
+n
) (-1)^a \binom{n}{a}}{n}\sum_{k=0}^a \binom{n}{k}
+\frac{(-1)^a}{2}\Bigg(
        \sum_{k=0}^a \binom{n}{k}\Bigg)^2\\
&-\frac{1}{2} 
\sum_{k=0}^a (-1)^k \binom{n}{k}^2
+\frac{1}{n}\sum_{k=0}^a (-1)^k k \binom{n}{k}^2,
\\
\sum_{k=0}^a (-1)^k \Big(
        \sum_{j=0}^k (-1)^j \binom{n}{j}\Big)^2=&\frac{1}{2} 
\sum_{k=0}^a (-1)^k \binom{n}{k}^2\\
&-\frac{
1}{n}\sum_{k=0}^a (-1)^k k \binom{n}{k}^2
+\frac{(-1)^a \binom{n}{a}^2 (-a
+n
)^2}{2 n^2},\\
\sum_{k=0}^a (-1)^k \Big(
        \sum_{j=0}^k \frac{1}{\binom{n}{j}}\Big)^2=&\, 
        \frac{(a+1)(-1)^a}{(n+2) \binom{n}{a}}\sum_{k=0}^a \frac{1}{\binom{n}{k}}
        +\frac{(-1)^a}{2} \Bigg(
                \sum_{k=0}^a \frac{1}{\binom{n}{k}}\Bigg)^2
\\
&+\frac{n}{2 (n+2)}\sum_{k=0}^a \frac{(-1)^k}{\binom{n}{k}^2}
-\frac{1}{n+2}\sum_{k=0}^a \frac{(-1)^k k}{\binom{n}{k}^2},\\
\sum_{k=0}^a (-1)^k \Big(
        \sum_{j=0}^k \frac{(-1)^j}{\binom{n}{j}}\Big)^2=&\,-\frac{n}{2 (n+2)}\sum_{k=0}^a \frac{(-1)^k}{\binom{n}{k}^2}
+\frac{1}{n+2}\sum_{k=0}^a \frac{(-1)^k k}{\binom{n}{k}^2}\\
&+\frac{n+1}{n+2}\sum_{k=0}^a \frac{1}{\binom{n}{k}}+\frac{(a+1) (n+1)}{(n+2)^2 \binom{n}{a}}\\
&+
        \frac{(n+1)^2(-1)^a}{2 (n+2)^2}
        +\frac{(a+1)^2(-1)^a}{2 (n+2)^2 \binom{n}{a}^2},
\end{align*}
where the first two identities are valid for $a,n\in\ZZ$ and $n\neq0$ and the last two identities are valid for $a,n\in\ZZ$ with $a\leq n$.

\section{Conclusion}\label{Sec:Conclusion}

In this article, under the umbrella of algorithmic symbolic summation, we established new algebraic connections between summation problems involving generic sequences and difference field/ring theory taking special care of concrete sequences arising in contexts like analysis, combinatorics, number theory and special functions. We feel this is only the ``first word'' in view of the high potential for applications of various kinds. One future application domain is summation identities involving elliptic functions or modular forms. This will be especially interesting in upcoming calculations~\cite{Elliptic:18} emerging in renormalizable Quantum Field Theories.
Another more concrete application domain is the area of $q$-identities involving $q$-hypergeometric series and sums. But already for $q=1$ one can study aspects of \textit{definite} summation. 
We plan to investigate these questions in forthcoming articles. For example, if we specialize our sums to definite versions by setting $a=n$ (and possibly consider the even or odd case), further simplifications can be achieved by \texttt{Sigma}. Typical examples are
\begin{align*}
\sum_{k=0}^n \Bigg(
        \sum_{j=0}^k \frac{1}{\binom{n}{j}}\Bigg)^2=&\,\frac{3 (n+1)^3 (n+2)}{4 (2 n+1) (2 n+3) \binom{2 n}{n}}\sum_{k=1}^n \frac{\binom{2 k}{k}}{k}+2^{-n-1} (n+1) 
\sum_{k=1}^n \frac{2^k}{k}\\
&
+2^{-2 n-3} (n+1)^2 (n+2) \Bigg(
        \sum_{k=1}^n \frac{2^k}{k}\Bigg)^2
+\frac{n^2+6 n+6}{2 (2 n+3)},\\
\sum_{k=0}^{2 n} (-1)^k \Bigg(
        \sum_{j=0}^k \frac{1}{\binom{2 n}{j}}\Bigg)^2=&\,
\frac{2^{-2 n-2} (2 n+1) (4 n+3)}{n+1}\sum_{k=1}^{2n}\frac{2^k}k\\
&+2^{-4 n-3} (2 n+1)^2 \Bigg(\sum_{k=1}^{2n}\frac{2^k}k\Bigg)^2
+\frac{3 n+2}{2 (n+1)},\\
\sum_{k=0}^{2n} (-1)^k \Bigg(
        \sum_{j=0}^k \binom{2 n}{j}\Bigg)^2=&\,2^{4 n-1},
\end{align*}
where the first two identities are valid for $n\geq0$ and the last identity holds for $n\geq1$.

\section{Appendix: Simple \rpisiSE-rings and algorithmic properties}\label{Sec:Appendix}

For a given difference ring (resp.\ field) $\dfield{\AA}{\sigma}$, i.e., a ring (resp.\ field) $\AA$ equipped with a ring (resp.\ field) automorphism $\fct{\sigma}{\AA}{\AA}$ the set of constants 
$\KK:=\const{\AA}{\sigma}=\{c\in\AA|\,\sigma(c)=c\}$ forms a subring (resp.\ subfield) of $\AA$. In this article we suppose that $\AA$ contains the rational numbers $\QQ$ as a subfield. Since $\sigma(1)=1$, this implies that $\QQ\subseteq\KK$ always holds. Moreover, by construction we will take care that $\KK$ will be always a field which will be called the constant field of $\dfield{\AA}{\sigma}$.

In the following we introduce the class of simple \rpisiSE-rings that forms the fundament of \texttt{Sigma}'s difference ring engine. Depending on the given input problem, the ground field is chosen accordingly among one of the following three difference fields.

\begin{psdefinition}\label{Def:BaseCaseFields}
We consider the following three difference fields $\dfield{\FF}{\sigma}$ with constant field $\KK$.
\begin{enumerate}
 \item[(1)] The \emph{rational case}: $\FF=\KK(k)$ where $\KK(k)$ is a rational function field and $\sigma(k)=k+1$.
 \item[(2)] The \emph{$q$-rational case}: $\FF=\KK(z)$ where $\KK(z)$ is a rational function field, $\KK=\KK'(q)$ is a rational function field ($\KK'$ is a field) and $\sigma(z)=q\,z$.
 \item[(3)] The \emph{mixed case}: $\dfield{\KK(k)(z_1,\dots,z_v)}{\sigma}$ where $\KK(k)(z_1,\dots,z_v)$ is a rational function field, $\KK=\KK'(q_1,\dots,q_v)$ is a rational function field ($\KK'$ is a field), $\sigma(k)=k+1$, and $\sigma(z_i)=q_i\,z_i$ for $1\leq i\leq v$.
\end{enumerate}
\end{psdefinition}

\noindent We remark that these difference fields can be embedded into the ring of sequences $\dfield{\seqK}{\Shift}$ as expected. For the rational case see Example~\ref{Exp:RatDR}, and for the other two cases we refer to~\cite[Ex.~5.3]{DR3}. Further aspects can be found in~\cite{Bauer:99}.

On top of such a ground field, a tower of extensions is built recursively depending on the input that is passed to \texttt{Sigma}. Let $\dfield{\AA}{\sigma}$ be the already constructed difference ring with constant field $\KK$. 
Then the tower can be extended by one of the following three types of extensions~\cite{Karr:81,DR1}; compare Definition~\ref{Def:SigmaExt}.
\begin{enumerate}
 \item[(1)]\textbf{\sigmaSE-extension:} Given $\beta\in\AA$, take the polynomial ring $\AA[t]$ ($t$ is transcendental over $\AA$) and extend the automorphism $\sigma$ from $\AA$ to $\AA[t]$ subject to the relation $\sigma(t)=t+\beta$. If $\const{\AA[t]}{\sigma}=\const{\AA}{\sigma}$, the difference ring $\dfield{\AA[t]}{\sigma}$ is called a \emph{\sigmaSE-extension} of $\dfield{\AA}{\sigma}$.
 \item[(2)]\textbf{\piE-extension:} Given a unit $\alpha\in\AA^*$, take the Laurent polynomial ring $\AA[t,t^{-1}]$ ($t$ is transcendental over $\AA$) and extend the automorphism $\sigma$ from $\AA$ to $\AA[t,t^{-1}]$ subject to the relation $\sigma(t)=\alpha\,t$ (and $\sigma(t^{-1})=\frac1{\alpha}\,t^{-1}$). If $\const{\AA[t,t^{-1}]}{\sigma}=\const{\AA}{\sigma}$, the difference ring $\dfield{\AA[t,t^{-1}]}{\sigma}$ is called a \emph{\piE-extension} of $\dfield{\AA}{\sigma}$.
 \item[(3)]\textbf{\rE-extension:} Given a primitive $\lambda$th root of unity $\alpha\in\KK$ with $\lambda\geq2$, take the algebraic ring $\AA[t]$ subject to the relation $t^{\lambda}=1$ and extend the automorphism $\sigma$ from $\AA$ to $\AA[t]$ subject to the relation $\sigma(t)=\alpha\,t$. If $\const{\AA[t]}{\sigma}=\const{\AA}{\sigma}$, the difference ring $\dfield{\AA[t]}{\sigma}$ is called an \emph{\rE-extension} of $\dfield{\AA}{\sigma}$.
\end{enumerate}
More generally, we call a difference ring $\dfield{\EE}{\sigma}$ an \emph{\rpisiSE-extension} of a difference ring $\dfield{\AA}{\sigma}$ if it is built by a tower 
\begin{equation}\label{Equ:ETower}
\AA=\EE_0\leq\EE_1\leq\dots\leq\EE_e=\EE
\end{equation}
of \rE-, \piE, and \sigmaSE-extensions starting from the difference ring $\dfield{\AA}{\sigma}$. Note that by construction we have that $\const{\EE}{\sigma}=\const{\AA}{\sigma}=\KK$. Finally, we restrict to the following case that is relevant for this article.

\begin{psdefinition}\label{Def:SimpleRing}
We call a difference ring $\dfield{\EE}{\sigma}$ a \emph{simple \rpisiSE-ring with constant field $\KK$} if it is an \rpisiSE-extension of a difference ring $\dfield{\AA}{\sigma}$ built by the tower~\eqref{Equ:ETower} with the following properties:
\begin{enumerate}
\item[(1)] $\dfield{\AA}{\sigma}$ is one of the three difference fields from Definition~\ref{Def:BaseCaseFields}; 
\item[(2)] for $i$ with $1\leq i\leq e$ the following holds: if $\dfield{\EE_i}{\sigma}$ is a \piE-extension of $\dfield{\EE_{i-1}}{\sigma}$ with $\EE_i=\EE_{i-1}[t_i,t_i^{-1}]$, then $\sigma(t_i)/t_i\in\AA^*$.
\end{enumerate}
\end{psdefinition}

Note that within such a simple \rpisiSE-ring the generators of 
\begin{enumerate}
\item[(a)] \rE-extensions model algebraic products of the form $\alpha^k$ where $\alpha$ is a primitive root of unity; 
\item[(b)] \piE-extensions model \hbox{(q--)}hypergeometric/mixed hypergeometric products depending on the chosen base field $\dfield{\AA}{\sigma}$;
\item[(c)] \sigmaSE-extensions represent nested sums whose summands are built recursively by polynomial expressions in terms of objects that are introduced in (a), (b) and (c).
\end{enumerate}

Given such a simple \rpisiSE-ring with constant field $\KK$, we can exploit the algorithmic properties summarized in Theorem~\ref{Thm:EmbedSigmaExt} that are incorporated within the summation package \texttt{Sigma}. For a detailed description of parts~(1) and~(2a) of Theorem~\ref{Thm:EmbedSigmaExt} we refer to~\cite[Section~7.2]{DR3}; for part~(2b) of Theorem~\ref{Thm:EmbedSigmaExt} we refer to~\cite[Section~5]{DR3}. 

In the following we sketch some further aspects.
Namely, given an expression $X(k)(=X_k)$ in terms of nested sums over hypergeometric (resp.\ $q$-hyper\-geometric or mixed hypergeometric) products, one can always construct algorithmically an \rpisiSE-ring $\dfield{\EE}{\sigma}$ together with an evaluation function $\fct{\ev}{\EE\times\NN}{\KK}$ with the following two properties (A) and (B).\\
(A) $\dfield{\EE}{\sigma}$ is constructed explicitly by the tower of extensions~\eqref{Equ:ETower} with the generators $t_i$ ($\EE_i=\EE_{i-1}[t_i]$ for \rE- or \sigmaSE-extensions and $\EE_i=\EE_{i-1}[t_i,t_i^{-1}]$ for a \piE-extension) where for $1\leq i\leq e$,
there is an explicitly given product or a nested sum over products, say $F_i(k)$, and a $\lambda_i\in\NN$ such that
$\ev(t_i,k)=F_i(k)$
holds for all $k\geq\lambda_i$. In particular, 
the resulting map $\fct{\tau}{\EE}{\seqK}$ with $\tau(f)\equiv(\ev(f,k))_{k\geq0}$ yields a $\KK$-embedding. 

\begin{psexample}
Consider the \rpisiSE-ring $\dfield{\KK(k)[s]}{\sigma}$ from Example~\ref{Exp:K(k)[s]}. There we obtained $\ev$ with $\ev(s,k)=H_k$ for all $k\geq\lambda$ with $\lambda=0$.
\end{psexample}

\noindent (B) One can construct an element $x\in\EE$ and a $\lambda\in\NN$ such that 
$X(i)=\ev(x,i)$ holds for all $i\geq\lambda$. 
In particular, this $x\in\EE$ can be rephrased again as an expression in terms of products or sums defined over such products in the following way: replacing the generators $t_i$ in $f$ by the attached sums or products\footnote{In the $q$-case (resp.\ in the mixed case) we also have to replace $z$ by $q^k$ (resp.\ $z_i$ by $q_i^k$ for $1\leq i\leq v$).} one gets an expression 
$X'(k)$ in terms of nested sums over products such that 
$X(k)=\ev(x,k)=X'(k)$ holds for all $k\in\NN$ with $k\geq\lambda$. 

In addition, the summation paradigms of refined parameterized telescoping~\cite{FastAlgorithm1,FastAlgorithm2,FastAlgorithm3,DR1,DR2,DR3} and recurrence solving can be carried out in such simple \rpisiSE-rings. In a nutshell, we can solve the telescoping problem and enhanced versions of it in the \rpisiSE-ring $\dfield{\EE}{\sigma}$ or equivalently in the product-sum sequence ring $\dfield{\SubS}{\Shift}$. This enables one to discover, e.g., the identities given in Section~\ref{Sec:Conclusion}. 

Furthermore, the difference ring algorithms combined with the algorithms given in~\cite{KS:06a} work also for difference rings where one starts with the free difference field $\dfield{\GG}{\sigma}$ introduced in Subsection~\ref{Sec:SymbolicApproach} as base field, adjoins the generators given in Definition~\ref{Def:BaseCaseFields}, and puts a tower of \rpisiSE-extensions on top; compare Subsection~\ref{Sec:SymbolicApproach}.

\medskip

\noindent\textbf{Acknowledgments.} We would like to thank Christian Krattenthaler for inspiring discussions. Special thanks go to Bill Chen and his collaborators at the center for Applied Mathematics at the Tianjin University for overwhelming hospitality in the endspurt phase of writing up this paper. We are especially grateful for all the valuable and detailed suggestions of the referee that improved substantially the quality of this article.


\begin{thebibliography}{10}

\bibitem{Elliptic:18}
J.~Ablinger, J.~Bl\"umlein, A.~De Freitas, M.~van Hoeij, E.~Imamoglu, C.G.
  Raab, C.S. Radu, and C.~Schneider.
\newblock Iterated elliptic and hypergeometric integrals for {F}eynman
  diagrams.
\newblock {\em J. Math. Phys.}, 59(062305):1--55, 2018.

\bibitem{AS:15}
J.~Ablinger and C.~Schneider.
\newblock Algebraic independence of sequences generated by (cyclotomic)
  harmonic sums.
\newblock {\em Annals of Combinatorics}, 22(2):213--244, 2018.

\bibitem{Abramov:71}
S.~A. Abramov.
\newblock On the summation of rational functions.
\newblock {\em Zh. Vychisl. Mat. Mat. Fiz.}, 11:1071--1074, 1971.

\bibitem{Abramov:89a}
S.~A. Abramov.
\newblock Rational solutions of linear differential and difference equations
  with polynomial coefficients.
\newblock {\em U.S.S.R. Comput. Math. Math. Phys.}, 29(6):7--12, 1989.

\bibitem{PAIV}
G.E. Andrews and P.~Paule.
\newblock Mac{M}ahon's partition analysis. {IV}. {H}ypergeometric multisums.
\newblock {\em S\'em. Lothar. Combin.}, 42:Art. B42i, 24, 1999.
\newblock The Andrews Festschrift (Maratea, 1998).

\bibitem{Bauer:99}
A.~Bauer and M.~Petkov{\v{s}}ek.
\newblock Multibasic and mixed hypergeometric {Gosper}-type algorithms.
\newblock {\em J.~Symbolic Comput.}, 28(4--5):711--736, 1999.

\bibitem{Gosper:78}
R.~W. Gosper.
\newblock Decision procedures for indefinite hypergeometric summation.
\newblock {\em Proc. Nat. Acad. Sci. U.S.A.}, 75:40--42, 1978.

\bibitem{Karr:81}
M.~Karr.
\newblock Summation in finite terms.
\newblock {\em J.~ACM}, 28:305--350, 1981.

\bibitem{Karr:85}
M.~Karr.
\newblock Theory of summation in finite terms.
\newblock {\em J.~Symbolic Comput.}, 1:303--315, 1985.

\bibitem{KS:06b}
M.~Kauers and C.~Schneider.
\newblock Application of unspecified sequences in symbolic summation.
\newblock In J.G. Dumas, editor, {\em Proc. ISSAC'06.}, pages 177--183. ACM
  Press, 2006.

\bibitem{KS:06a}
M.~Kauers and C.~Schneider.
\newblock Indefinite summation with unspecified summands.
\newblock {\em Discrete Math.}, 306(17):2021--2140, 2006.

\bibitem{DR4}
E.D. Ocansey and C.~Schneider.
\newblock Representing (q-)hypergeometric products and mixed versions in
  difference rings.
\newblock In C.~Schneider and E.~Zima, editors, {\em { Advances in Computer
  Algebra. WWCA 2016.}}, volume 226 of {\em Springer Proceedings in Mathematics
  \& Statistics}, pages 175--213. Springer, 2018.
\newblock arXiv:1705.01368 [cs.SC].

\bibitem{Paule:95}
P.~Paule and M.~Schorn.
\newblock A {M}athematica version of {Z}eilberger's algorithm for proving
  binomial coefficient identities.
\newblock {\em J.~Symbolic Comput.}, 20(5--6), 1995.

\bibitem{AequalB}
M.~Petkov{\v s}ek, H.~S. Wilf, and D.~Zeilberger.
\newblock {\em $A=B$}.
\newblock A K Peters, Wellesley, MA, 1996.

\bibitem{Schneider:07d}
C.~Schneider.
\newblock Simplifying sums in {$\Pi\Sigma$}-extensions.
\newblock {\em J. Algebra Appl.}, 6(3):415--441, 2007.

\bibitem{Sigma1}
C.~Schneider.
\newblock Symbolic summation assists combinatorics.
\newblock {\em Sem.~Lothar. Combin.}, 56:1--36, 2007.
\newblock Article B56b.

\bibitem{FastAlgorithm3}
C.~Schneider.
\newblock A refined difference field theory for symbolic summation.
\newblock {\em J. Symbolic Comput.}, 43(9):611--644, 2008.
\newblock arXiv:0808.2543 [cs.SC].

\bibitem{FastAlgorithm1}
C.~Schneider.
\newblock Structural theorems for symbolic summation.
\newblock {\em Appl. Algebra Engrg. Comm. Comput.}, 21(1):1--32, 2010.

\bibitem{DR2}
C.~Schneider.
\newblock A streamlined difference ring theory: Indefinite nested sums, the
  alternating sign and the parameterized telescoping problem.
\newblock In F.~Winkler, V.~Negru, T.~Ida, T.~Jebelean, D.~Petcu, S.~Watt, and
  D.~Zaharie, editors, {\em Symbolic and Numeric Algorithms for Scientific
  Computing (SYNASC), 2014 15th International Symposium}, pages 26--33. IEEE
  Computer Society, 2014.
\newblock arXiv:1412.2782v1 [cs.SC].

\bibitem{FastAlgorithm2}
C.~Schneider.
\newblock Fast algorithms for refined parameterized telescoping in difference
  fields.
\newblock In J.~Gutierrez, J.~Schicho, and M.~Weimann, editors, {\em Computer
  Algebra and Polynomials, Applications of Algebra and Number Theory}, volume
  8942 of {\em Lecture Notes in Computer Science (LNCS)}, pages 157--191.
  Springer, 2015.
\newblock arXiv:1307.7887 [cs.SC].

\bibitem{DR1}
C.~Schneider.
\newblock A difference ring theory for symbolic summation.
\newblock {\em J. Symb. Comput.}, 72:82--127, 2016.
\newblock arXiv:1408.2776 [cs.SC].

\bibitem{DR3}
C.~Schneider.
\newblock Summation theory {II}: Characterizations of {$R\Pi\Sigma$}-extensions
  and algorithmic aspects.
\newblock {\em J. Symb. Comput.}, 80(3):616--664, 2017.
\newblock arXiv:1603.04285 [cs.SC].

\bibitem{Singer:97}
M.~van~der Put and M.F. Singer.
\newblock {\em Galois theory of difference equations}, volume 1666 of {\em
  Lecture Notes in Mathematics}.
\newblock Springer-Verlag, Berlin, 1997.

\bibitem{Zeilberger:90b}
D.~Zeilberger.
\newblock A fast algorithm for proving terminating hypergeometric identities.
\newblock {\em Discrete Math}, 80(2):207--211, 1990.

\end{thebibliography}

\end{document}